\documentclass[11pt]{article}

\usepackage{amsmath,amssymb}
\usepackage{cite}
\usepackage[dvips]{color,graphicx}
\begin{document}

\title{\Large\bf Multi-antikaonic nuclei in the relativistic mean-field theory }

\author{Takumi Muto\thanks{Email address: takumi.muto@it-chiba.ac.jp} \\
{\it Department of Physics, Chiba Institute of Technology,
2-1-1 Shibazono} \\
{\it Narashino, Chiba 275-0023, Japan} \\
 Toshiki Maruyama\thanks{Email address: maruyama.toshiki@jaea.go.jp}\\
 {\it Advanced Science Research Center, Japan Atomic Energy Agency} \\
{\it Tokai, Ibaraki 319-1195, Japan} \\
Toshitaka Tatsumi\thanks{Email address: tatsumi@ruby.scphys.kyoto-u.ac.jp}\\
{\it Department of Physics, Kyoto University, 
Kyoto 606-8502, Japan}
}

\date{\today}
\maketitle

\vspace{1.0cm}
      
\begin{abstract}
Properties of multi-antikaonic nuclei (MKN), where several numbers of $K^-$ mesons are bound, are studied in the relativistic mean-field model, combined with chiral dynamics for kaonic part of the thermodynamic potential. The density profiles for nucleons and $K^-$ mesons, the single particle energy of the $K^-$ mesons, and binding energy of the MKN are obtained. The effects of the $\bar K-\bar K$ interactions on these quantities are discussed in comparison with other meson ($\sigma$, $\omega$, and $\rho$)-exchange models. It is shown that the $\bar K-\bar K$ interactions originate from two contributions: One is the contact interaction between antikaons inherent in chiral symmetry, and the other is the one generated through coupling between the $K^-$ and meson mean fields. Both effects of the $\bar K-\bar K$ repulsive interactions become large on the ground state properties of the MKN as the number of the embedded $K^-$  mesons increases. A relation between the multi-antikaonic nuclei and kaon condensation in infinite and uniform matter is mentioned.
\end{abstract}

\begin{description} 
{\footnotesize\item PACS: 05.30.Jp, 11.30.Rd, 21.85.+d, 26.60.-c.  
\item Keywords:  multi-antikaonic nuclei; chiral symmetry; relativistic mean-field theory}
\end{description}

\newpage
\section{Introduction}
\label{sec:intro}

\ \ Strangeness and chiral symmetry play important roles
on understanding hadron dynamics in highly dense matter.
In neutron-star matter, kaon condensation, as macroscopic appearance of strangeness, has been widely studied theoretically and
observationally\cite{kn86,mtt93,tpl94,lbm95,fmmt96,pbp97}. The existence of kaon-condensed phase in neutron stars would soften the hadronic equation of state (EOS) and modify bulk properties of neutron stars such as the mass-radius relation\cite{tpl94,lbm95,fmmt96,pbp97}. Effects of the phase-equilibrium condition associated with the first-order phase transition 
on the nonuniform structure of neutron stars have been
investigated\cite{g01,cgs00,nr01,vyt03,mtv06,mtec06}. Delayed collapse of protoneutron stars accompanying a phase transition to the kaon-condensed phase has also been discussed\cite{bb94,ty99,p00,p01}. Cooling of neutron stars would be largely accelerated through rapid neutrino emission mechanisms in the presence of kaon condensates\cite{bk88,t88,pb90,fmtt94}. 
It depends on in-medium kaon properties whether kaon condensation is realized in dense matter or not. To extract information on the EOS and kaon and antikaon properties in a dense medium, experiments on kaon and antikaon production in heavy-ion collisions have been performed with the kaon spectrometer (KaoS) at GSI\cite{s08}. Effects of nuclear correlations on kaon condensation have been theoretically elucidated\cite{ppt95,wrw97,chp00}. 

It has also been suggested that hyperons ($\Lambda$, $\Sigma$, $\Xi$, $\cdots$) may be mixed at several times of the standard nuclear matter density in neutron stars\cite{g85,gm91,h98,phz99,h00,bbs98,v00,y02,t04}. The mixing of hyperons may as well cause softening of the EOS and resulting dynamical and thermal evolutions of neutron stars\cite{h98,p99,t04}. 
The interplay between kaon condensation and hyperons and their possible coexistence in neutron stars have been elucidated\cite{m93,kvk95,ekp95,kpe95,sm96,pbg00,bb01,m02,kv03,mpp05,rhhk07,m08a}. 

As cold and dense hadronic states with strangeness which may be formed in laboratory experiments, deeply bound kaonic nuclear states have been proposed based on strongly attractive $\bar K N$ interaction\cite{ay02}. Experimental proposals for creating  double and multiple kaon clusters have also been made\cite{yda04}. 
Recently, possibilities of bound nuclear systems with several antikaons such as multi-antikaonic nuclei\cite{gfgm07,mmt07} and a $\bar K\bar KN$ molecular state\cite{ej08}
have been considered theoretically.  
Possible existence of the deeply bound kaonic nuclei has not yet been confirmed experimentally\cite{i03,k05,a05}, but the relevant kaon-baryon interaction in a nuclear medium has been discussed by the use of both microscopic models and phenomenological models\cite{gh08}. 

There is a qualitative difference in formation mechanisms between kaon condensation in neutron stars and kaonic nuclei in terrestrial experiments. The kaon-condensed state is a strangeness-nonconserving system, and $K^-$ mesons are created spontaneously from normal matter through the weak interaction processes such as $N + n \rightarrow N + p + K^-$, $N + e^- \rightarrow N + K^- + \nu_e$ ($N=p,n$) beyond a critical density where the lowest energy for $K^-$, $\omega_{K^-}$, meets the charge chemical potential $\mu$ (=$\mu_{K^-}=\mu_e$). The ground state of the $K^-$-condensed phase is then in chemical equilibrium for the weak processes, $\mu_n=\mu_p+\mu_{K^-}$, $\mu_e=\mu_{K^-}$\cite{mt92,mti00}. On the other hand, kaonic nuclear state is a strangeness-conserving system since  their formation occurs in a strong interaction time scale, and  strangeness should be implanted by trapping of $K^-$ mesons inside nuclei. The channel-couplings,  $\bar K N\rightleftharpoons \pi Y$, $\bar KNN\rightleftharpoons YN$ ($Y=\Lambda, \Sigma$), etc., may also be relevant to formation and structure of kaonic nuclei, whose decay modes are given by these strong processes\cite{gh08}. 
In spite of such a difference, kaon condensation and kaonic nuclei show qualitatively common features.  As an instance, in $K^-$-condensed state, proton fraction becomes as large as neutrons, and electrons become less dominant, since the negative charge carried by $K^-$ condensates is compensated as a result of the increase in the positive charge given by the protons and since negative charge of the electrons are taken over by that of $K^-$ condensates\cite{mtt93,tpl94}. Thus the particle composition of the $K^-$-condensed state is similar to that of kaonic nuclei. 

Both kaon condensation and kaonic nuclei may be closely related with each other in the context of kaon dynamics in hadronic matter.     
In this paper, we consider the properties of multi-antikaonic nuclei (abbreviated as MKN) where several antikaons ($K^-$ mesons) are bound in the ground state of the nucleus.
 The density profiles for nucleons and $K^-$ mesons, the single particle energy of the $K^-$ mesons, and binding energy of the MKN are obtained. 
It is clarified how the structure of highly dense and deeply bound meson-baryon system with multi-strangeness is formed with increase in the number of the embedded $K^-$ mesons. With this systematic change of strangeness, we address possible continuity between the MKN as a finite system and uniform kaon condensation in infinite matter which may be realized in neutron stars.\footnote{Preliminary results on the MKN within our framework have been reported in Ref.~\cite{mmt07}.}
 
The thermodynamic potential is given in the relativistic mean-field
theory (RMF) 
with reference to the density functional theory, which has been applied to study the nonuniform structure of kaon-condensed phase in neutron-star matter\cite{mtv06,mtec06}. 
In this paper, we base our model on the RMF with extension to include
kaon dynamics that respects chiral symmetry\cite{mmt07}, while the
meson-exchange model has been often applied as a phenomenological model to describe kaon condensation
in neutron stars\cite{g01,cgs00,nr01,vyt03,mtv06,mtec06,p00,p01,kpe95,sm96,bb01},
kaonic pastas \cite{g01,cgs00,nr01,vyt03,mtv06,mtec06} and recently the kaonic nuclei\cite{mfg05,zpln06,gfgm07}. In the context of kaon condensation in dense matter, it has been recognized that chiral symmetry is essential and has many implications on the properties of the condensed phase\cite{kn86,mtt93,tpl94,lbm95,fmmt96,pbp97} 
as well as the onset mechanism\cite{mt92,mti00}. The most important difference between our model presented in this paper and the meson-exchange models lies in the nonlinearities of the kaon field dictated by chiral symmetry.  
Thus, we can take into account the nonlinear effects in the MKN,  
the leading effect of which may be the repulsive interaction among
kaons \cite{mmt07}. 

We concentrate on the effects of the $\bar K-\bar K$ interactions on the ground state properties of the MKN. The $\bar K-\bar K$ interactions are classified into two contributions in our model: One is the contact interaction between antikaons inherent in chiral symmetry, where the nonlinear terms originating from the nonlinear representation of the $\bar K$ field are incorporated in the chiral Lagrangian, and the other is the one generated through couplings  between the $K^-$ and meson ($\sigma$, $\omega$, and $\rho$) mean fields. The latter is a common contribution also incorporated in  other meson-exchange models. 
It is emphasized that both parts of the $\bar K-\bar K$ interactions have sizable contributions to the properties of the MKN as the number of the embedded $K^-$ mesons increases.    

The paper is organized as follows. Section~\ref{sec:formulation} gives the formulation to obtain the gross structure of the MKN. In 
Sec.~\ref{sec:results}, numerical results and discussion are given. Summary and concluding remarks are devoted in Sec.~\ref{sec:summary}. 

\section{Formulation}
\label{sec:formulation}

\subsection{Chiral model}
\label{subsec:chiral}

\ \ The MKN is described as a nucleus where several $K^-$ mesons are bound through the attractive $\bar K N$ interaction. We assume a spherical symmetry for the MKN, and the density profiles and the other quantities are given as functions of the radial distance $r$ from the center of the MKN. The number of the embedded $K^-$ mesons with the lowest energy $\omega_{K^-}$ is denoted as $|S|$, and  the mass number and atomic number of the initial target nucleus as $A$ and $Z$, respectively.  
In this paper we consider such kaons by postulating that they are
condensed in the lowest energy state and treat them within the
mean-field approximation.\footnote{Note that we can only fix the average number of kaons in this treatment 
except one kaon case.} 
The classical $K^-$ field is represented as 
\begin{equation}
K^-(r)=\frac{f}{\sqrt{2}}\theta(r) \ , 
\label{eq:kfield}
\end{equation}
where $\theta(r)$  is the chiral angle and $f$ (= 93 MeV) is the meson decay constant. 
Kaon dynamics including the $s$-wave $\bar KN$ interaction is incorporated through the kaonic part of the thermodynamic potential ($\Omega_K$) that is obtained from the nonlinear chiral Lagrangian\cite{ty99}. 
 The thermodynamic potential $\Omega$ is then separated into contributions from the nucleons, $K^-$, other mesons ($\sigma$, $\omega$, $\rho$)\footnote{
For simplicity, we do not take into account the $\phi$ and $\sigma^\ast$ mesons which mediate a part of $\bar K-\bar K$ interactions. In Ref.~\cite{gfgm07}, these mesons have been shown to have minor roles on the $\bar K-\bar K$ interactions. }
 , and the Coulomb potential $V_{\rm Coul}(r)$:
$\Omega=\Omega_N+\Omega_K+\Omega_M +\Omega_{\rm Coul} $ with 
\begin{subequations}\label{eq:tpot}
\begin{eqnarray}
\Omega_N &=&\sum_{a=p,n}\int d^3 r\Bigg\lbrack\int_0^{k_{F,a}}\frac{d^3 k}{4\pi^3}\sqrt{m_N^{\ast 2}+k^2}-\rho_a\nu_a \Bigg\rbrack  \label{eq:tpotn} \\
\Omega_K &=&\int d^3 r \Bigg\lbrack f^2 (1-\cos\theta)\left(m_K^{\ast 2}-2\widetilde\omega_{K^-} X_0\right) 
-\frac{1}{2}\widetilde\omega_{K^-}^2 f^2\sin^2\theta+\frac{f^2}{2}(\nabla\theta)^2\Bigg\rbrack \label{eq:tpotk} \\
\Omega_M &=&\int d^3 r\Bigg\lbrack\frac{1}{2}(\nabla\sigma)^2+\frac{1}{2}m_\sigma^2\sigma^2+U(\sigma) \cr
&-&\frac{1}{2}(\nabla\omega_0)^2-\frac{1}{2}m_\omega^2\omega_0^2-\frac{1}{2}(\nabla R_0)^2-\frac{1}{2}m_\rho^2 R_0^2 \Bigg\rbrack \label{eq:tpotm} \\
\Omega_{\rm Coul} &=&- \int d^3 r (\nabla V_{\rm Coul})^2 /(8\pi e^2)  \ . \label{eq:tpotc}
\end{eqnarray}
\end{subequations}
The quantities in Eq.~(\ref{eq:tpot}) are defined in terms of the coupling constants $g_{iN}$ and $g_{iK}$ ($i$ = $\sigma$, $\omega$, $\rho$).\footnote{There is a discussion about  ambiguity of the $\sigma \bar K \bar K$ coupling schemes\cite{tko07}. In this paper, the $\sigma$ meson mediating the scalar $\bar K N$ interaction is introduced phenomenologically, as is always the case within the RMF model. } 
In Eq.~(\ref{eq:tpotn}), the effective nucleon mass is given by $m_N^\ast (r)=m_N-g_{\sigma N}\sigma(r)$, where $\sigma(r)$ is the $\sigma$-mean field. $\rho_a(r)$ ($a$ = $p, n$) is the nucleon number density, and $\nu_a(r)\equiv (m_N^{\ast}(r)^2+k_{F, a}^2)^{1/2}$  with $k_{F, a}$ being the Fermi momentum. It is to be noted that $\nu_a(r)$ is related by the nucleon chemical potential as
$\nu_p(r)= \mu_p+V_{\rm Coul}(r)-g_{\omega N}\omega_0(r)-g_{\rho N}R_0(r)$ and $\nu_n(r)=\mu_n-g_{\omega N}\omega_0(r)+g_{\rho N}R_0(r)$, where $\omega_0(r)$ and $R_0(r)$ are the time components of the vector mean fields. 
In Eq.~(\ref{eq:tpotk}), the effective kaon mass squared is given by
$m_K^{\ast 2}(r)=m_K^2-2g_{\sigma K}m_K\sigma({\bf
r})$,\footnote{The definition of $g_{\sigma K}$ in our paper is the same
as that in Refs.~\cite{g01,nr01,mtv06,mtec06} and is different from that
in Refs.~\cite{p01,gfgm07} by a factor of 2.} and $\widetilde\omega_{K^-}(r)$ is the $K^-$ energy shifted by the Coulomb potential, $\widetilde\omega_{K^-}(r)=\omega_{K^-}-V_{\rm Coul} (r)$. The function $X_0$ is defined as $X_0(r)=g_{\omega K}\omega_0(r)+g_{\rho K}R_0(r)$, and the term including $X_0(r)$ in Eq.~(\ref{eq:tpotk}) represents the $\bar KN$ vector interaction term. 
In Eq.~(\ref{eq:tpotm}), the scalar self-interaction potential $U(\sigma)$ is given by $U(\sigma)$=$bm_N(g_{\sigma N}\sigma)^3/3+c(g_{\sigma N}\sigma)^4/4$ with $b$=0.008659 and $c$=$-$0.002421\cite{gm91}.
It is to be noted that the nonlinear terms of $\bar K-\bar K$ interactions are fully incorporated in the thermodynamic potential $\Omega$ [Eq.~(\ref{eq:tpot})] through the nonlinear representation of the Nambu-Goldstone boson field within chiral symmetry\cite{w96}. 

The coupled equations for the $K^-$, $\sigma$, $\omega$, and $\rho$ mesons are derived from Eq.~(\ref{eq:tpot}) in the local density approximation for nucleons:
\begin{subequations}\label{eq:eom}
\begin{eqnarray}
 -\nabla^2\sigma+m_\sigma^2\sigma&=&-\frac{dU}{d\sigma}+g_{\sigma N}(\rho_p^s+\rho_n^s)+2g_{\sigma K}m_Kf^2(1-\cos\theta) \ ,  \label{eq:eom1}\\
 -\nabla^2\omega_0+m_\omega^2\omega_0&=&g_{\omega N}(\rho_p+\rho_n)-2g_{\omega K}\widetilde\omega_{K^-} f^2(1-\cos\theta) \ ,  \label{eq:eom2}\\
 -\nabla^2 R_0+m_\rho^2 R_0&=&g_{\rho N}(\rho_p-\rho_n)-2g_{\rho K} \widetilde\omega_{K^-} f^2(1-\cos\theta) \ , \label{eq:eom3}\\
 \nabla^2\theta&=&\sin\theta\left(m_K^{\ast 2}-2\widetilde\omega_{K^-} X_0-\widetilde\omega_{K^-}^2\cos\theta\right) \ ,  \label{eq:eom4}\\
\nabla^2 V_{\rm Coul}&=&4\pi e^2(\rho_p-\rho_{K^-}) \ , 
 \label{eq:eom5}
\end{eqnarray}
\end{subequations}
where $\rho_p^s(r)$ ($\rho_n^s(r)$) is the scalar density of the proton (the neutron), $\rho_p(r)$ ($\rho_n(r)$) the number density of the proton (the neutron), and $\rho_{K^-}(r)$ is the number density of the $K^-$ mesons defined by the negative strangeness number density,
\begin{equation}
\rho_{K^-}(r)=\widetilde\omega_{K^-}(r)f^2\sin^2\theta(r)+2X_0(r) f^2(1-\cos\theta(r)) \label{eq:strangeness} \ .
\end{equation} 
The coupled equations (\ref{eq:eom1})$-$(\ref{eq:eom5}) with Eq.~(\ref{eq:strangeness}) are solved self-consistently under the constraints that the numbers of protons $Z$, strangeness $S$, and nucleons $A$ are fixed:
\begin{equation}
\int d^3 r\rho_{p}(r)=Z \ , \int d^3 r\rho_{K^-}(r)=|S|, \ \int d^3 r \rho_{\rm B}(r)=A \ .
\label{eq:constraint}
\end{equation} 
In Eq.~(\ref{eq:constraint}), the number of the $K^-$ mesons $|S|$ is given as the expectation value of the negative strangeness number operator, since the classical $K^-$ field (\ref{eq:kfield}) is not the eigenstate of the $K^-$ number operator. 
Throughout this paper, we call the framework based on the thermodynamic potential (\ref{eq:tpot}) and the resulting equations of motion (\ref{eq:eom1})$-$(\ref{eq:eom5}) the {\it chiral model}. 

If one discards the derivative terms and the kaon-coupling terms including $g_{iK}$ ($i$ = $\sigma$, $\omega$, $\rho$) in Eqs.~(\ref{eq:eom1})$-$(\ref{eq:eom3}), and neglecting the scalar self-interaction term, $-dU/d\sigma$, in Eq.~(\ref{eq:eom1}), one obtains
\begin{equation}
\sigma=\frac{g_{\sigma N}}{m_\sigma^2}(\rho_p^s+\rho_n^s) \ , \ \omega_0=\frac{g_{\omega N}}{m_\omega^2}(\rho_p+\rho_n)\ , \ R_0=\frac{g_{\rho N}}{m_\rho^2}(\rho_p-\rho_n) \ .
\label{eq:eoma}
\end{equation}
If one further imposes the following relations among the coupling constants\cite{p00}, 
\begin{equation}
\frac{g_{\sigma N} g_{\sigma K}}{m_\sigma^2}=\frac{\Sigma_{KN}}{2m_K f^2} \ , \ \frac{g_{\omega N} g_{\omega K}}{m_\omega^2}=\frac{3}{8f^2} \ , \ \frac{g_{\rho N} g_{\rho K}}{m_\rho^2}=\frac{1}{8f^2} 
\label{eq:relation}
\end{equation}
with $\Sigma_{KN}$ being the $KN$ sigma term, one obtains 
$m_K^{\ast 2}(r)=m_K^2-\Sigma_{KN}(\rho_p^s+\rho_n^s)/f^2$ and $X_0=(\rho_p+\rho_n/2)/(2f^2)$, the latter of which represents the $K^--N$ vector interaction corresponding to the Tomozawa-Weinberg term.
These terms representing the $\bar K N$ scalar and vector interactions, respectively, are essentially equivalent to those derived on the basis of chiral symmetry by the use of the nonlinear chiral Lagrangian\cite{kn86,ty99,ekp95}. 

In this paper, the coupling constants are chosen as follows: $g_{i N}$ ($i$ = $\sigma$, $\omega$, $\rho$) are determined to reproduce saturation properties of nuclear matter, binding energies and proton-mixing ratios, and density distributions of protons and neutrons in finite nuclei\cite{mtv06,mtec06}. 
The vector meson-$\bar K$ coupling constants are set to be $g_{\omega K} =g_{\omega N}/3$=2.91, $g_{\rho K} = g_{\rho N}$=4.27 from the quark and isospin counting rule\cite{g01,mtv06,mtec06}.\footnote{
 The vector meson dominance hypothesis may also give another constraint to the strengths of $g_{\omega K}$ and $g_{\rho K}$\cite{f72}. Here we follow the conventional choice of these parameters. }
The scalar meson-$\bar K$ coupling constant $g_{\sigma K}$ is related with the $K^-$ optical potential depth $U_K$ in symmetric nuclear matter at the standard nuclear density $\rho_0$ (=0.153 fm$^{-3}$) in the form, $U_K$=$-(g_{\sigma K}\langle\sigma\rangle+g_{\omega K}\langle\omega_0\rangle)$, where $\langle\sigma\rangle$ and $\langle\omega_0\rangle$ are meson mean-fields at $\rho_p=\rho_n=\rho_0/2$. For given $U_K$ and $g_{\omega K}$, the value of $g_{\sigma K}$ is fixed.   

The parameters used in our model are listed in Table~\ref{tab:para}. We consider two cases for $U_K$, $U_K$=$-$80 MeV and $-$ 120 MeV. From the first relation of Eq.~(\ref{eq:relation}), the $KN$ sigma term, $\Sigma_{KN}$, is estimated as $\Sigma_{KN}$= 332 MeV (754 MeV) for $U_K$=$-$80 MeV ($-$ 120 MeV). The case of $U_K$=$-$ 120 MeV should be regarded as an extreme case of strongly attractive $\bar K- N$ interaction.
It should be noted that our choice of the coupling constants, $g_{\omega K}$ and $g_{\rho K}$ in Table.~\ref{tab:para}, does not strictly meet the second and third relations in Eq.~(\ref{eq:relation}). Those satisfying these relations, denoted as $g_{\omega K}^{\rm T.W.}$ and $g_{\rho K}^{\rm T.W.}$, respectively, are given as $g_{\omega K}^{\rm T.W.}$=3.05 and $g_{\rho K}^{\rm T.W.}$=2.00 after substitution of the other parameters listed in Table~\ref{tab:para} 
in Eq.~(\ref{eq:relation}). Thus the value of $g_{\omega K}$ used in this paper is almost the same as that of $g_{\omega K}^{\rm T.W.}$,  whereas the value of $g_{\rho K}$ is bigger than $g_{\rho K}^{\rm T.W.}$ by a factor 2.  It is left as a future work to consider how the results rely on the choice of the coupling constants, especially $g_{\rho K}$. 

\begin{table}[!]
\caption{The parameters used in our model. The $K^-$ optical potential $U_K$ for the symmetric nuclear matter at the saturation density is given as $U_K$=$-(g_{\sigma K}\langle\sigma\rangle+g_{\omega K}\langle\omega_0\rangle)$, where $\langle\sigma\rangle$ and $\langle\omega_0\rangle$ are meson mean-fields at $\rho_p=\rho_n=\rho_0/2$.}
\begin{center}
\begin{tabular}{c c c c c c c c c c}
\hline
$g_{\sigma N}$ & $g_{\omega N}$ & $g_{\rho N}$ & $b$ & $c$ & $m_\sigma$ & $m_\omega$ & $m_\rho$ & $g_{\omega K}$ & $g_{\rho K}$  \\
    & & & && (MeV) & (MeV) & (MeV) \\\hline
6.39 & 8.72 & 4.27 & 0.008659 & $-$ 0.002421 & 400 & 783 & 769 & $g_{\omega N}/3$ & $g_{\rho N}$ \\\hline
\end{tabular}
\vspace{0.5cm}

\noindent
\begin{tabular}{c c}
\hline
$g_{\sigma K}$ & [$U_K$ (MeV)] \\
0.97 & $-$80       \\\hline
2.21 & $-$120    \\\hline
\end{tabular}
\label{tab:para}
\end{center}
\end{table}

It is to be noted that a pole contribution of the $\Lambda$(1405) (abbreviated as $\Lambda^\ast$) as well as the range terms proportional to $\omega_{K^-}^2$ in the $K^-$ self-energy is important to be empirically consistent with the $s$-wave $K$ ($\bar K$)-$N$ scattering amplitudes\cite{lbm95,fmmt96}. Furthermore, Akaishi and Yamazaki proposed deeply bound kaonic nuclear states based on the assumption that the $\Lambda^\ast$ is a bound state of the $K^-$ and proton\cite{ay02}.
Recently, the two-pole structure of the $\Lambda^\ast$ has been suggested, which makes the discussion on the structure of kaonic nuclei based more delicate\cite{mor05}.

In our framework concerning kaon condensation, the $\Lambda^\ast$ has been regarded as an elementary particle and the range terms have also been taken into account\cite{fmmt96}. It has been shown that these contributions to the energy density become negligible as far as $K^-$ condensation is concerned,  since the lowest $K^-$ energy $\omega_{K^-}$ (=$\mu_{K^-}$) becomes so small [$O(m_\pi)]$ and that sum of the $K^-$ and proton  energies lies well below the $\Lambda^\ast$ pole\cite{fmmt96}. 
However, as we see in Sec.~\ref{subsubsec:3-2-2}, the $\omega_{K^-}$ is not very small as compared with the free kaon mass $m_K$ for $U_K$=$-$80 MeV.
Throughout this paper, we discard these correction terms and consider the simplified expression for the energy of the MKN. The studies on the effects of the $\Lambda^\ast$ and the range terms are in progress and will be reported elsewhere. 

\subsection{Meson-exchange models}
\label{subsec:mem}

\ \ For the description of the in-medium $\bar K N$ interactions and kaon condensation in neutron stars, the meson-exchange models (MEM) have often been used\cite{kpe95,sm96,p00,p01,g01, cgs00,nr01,vyt03,mtv06,mtec06,gfgm07}. The MEMs have been shown to work well for low energy $\bar K N$ dynamics, although it does not respect chiral symmetry. They are two types of MEMs used in the literatures (we call them MEM1 and MEM2). We show, for comparison, that the two MEMs are derived from the chiral model obtained in Sec.~\ref{subsec:chiral} in the following prescriptions. The MEM1 is obtained by the expansion of the kaonic part of the thermodynamic potential $\Omega_K$ [ (\ref{eq:tpotk}) ] 
 in lowest order with respect to the $K^-$ field $\theta$ by the use of the approximations, $\sin\theta\rightarrow \theta$, $\cos\theta\rightarrow 1-\theta^2/2$, and by addition of  the two terms (i) $\displaystyle \int d^3 r (-\frac{f^2}{2}\theta^2X_0^2)$ and (ii) $\displaystyle\int d^3 r \frac{f^2}{2}\theta^2(g_{\sigma K}\sigma)^2$ to the $\Omega_K$.
 The resulting kaonic part of the thermodynamic potential, $\Omega_{K, {\rm MEM1}}$, is written as
\begin{equation}
 \Omega_{K,{\rm MEM1}}=\int d^3 r \Bigg\lbrack\frac{f^2}{2} \theta^2\Bigg\lbrace {m_K^\ast}_{\rm ,MEM}^2-(\widetilde\omega_{K^-}+X_0)^2\Bigg\rbrace +\frac{f^2}{2}(\nabla\theta)^2\Bigg\rbrack \ , 
 \label{eq:tpotkmem1}
 \end{equation}
 where ${m_K^\ast}_{\rm ,MEM1}\equiv m_{K^-}-g_{\sigma K}\sigma$. 
 The equations of motion for the MEM1 then read
\begin{subequations}\label{eq:lin}
\begin{eqnarray}
 -\nabla^2\sigma+m_\sigma^2\sigma&=&-\frac{dU}{d\sigma}+g_{\sigma N}(\rho_n^s+\rho_p^s)+g_{\sigma K}{m_K^\ast}_{\rm ,MEM1} f^2\theta^2 \ ,  \label{eq:lin1}\\
 -\nabla^2\omega_0+m_\omega^2\omega_0&=&g_{\omega N}(\rho_n+\rho_p)-g_{\omega K}(\widetilde\omega_{K^-}+X_0) f^2\theta^2 \ ,  \label{eq:lin2}\\
 -\nabla^2 R_0+m_\rho^2 R_0&=&g_{\rho N}(\rho_p-\rho_n)-g_{\rho K}(\widetilde\omega_{K^-}+X_0) f^2\theta^2 \ , \label{eq:lin3}\\
 \nabla^2\theta&=&\Big\lbrack{m_K^\ast}_{\rm ,MEM1}^2-(\widetilde\omega_{K^-}+X_0)^2\Big\rbrack\theta \ .\label{eq:lin4}\\
\nabla^2 V_{\rm Coul}&=&4\pi e^2(\rho_p-\rho_{K^-,{\rm MEM}}) \ , \label{eq:lin5}
\end{eqnarray}
\end{subequations}
where $\rho_{K^-,{\rm MEM}}=f^2\theta^2(\widetilde\omega_{K^-}+X_0)$. The term (i) stems from the minimal coupling between the kaon and the vector meson (abbreviated as $\bar K \bar K VV$) satisfying that the vector field should be coupled to a conserved current\cite{sm96}, and it works as an attractive contribution to the energy of the MKN. The term (ii) leads to another coupling between the kaon and the scalar field $\sigma$ (abbreviated as $\bar K \bar K \sigma\sigma$). It works as a repulsive contribution to the energy of the MKN. This $\bar K$-meson coupling scheme in the MEM1 is the same as that in Refs.~\cite{g01,cgs00,nr01,p01}. 

The MEM2 is the model where $\bar K\bar K\sigma\sigma$ term (ii) is omitted and only the term (i) is retained in the MEM1. 
 The resulting kaonic part of the thermodynamic potential, $\Omega_{K, {\rm MEM2}}$, is written as
\begin{equation}
 \Omega_{K,{\rm MEM2}}=\int d^3 r \Bigg\lbrack\frac{f^2}{2} \theta^2\Bigg\lbrace m_K^{\ast 2}-(\widetilde\omega_{K^-}+X_0)^2\Bigg\rbrace +\frac{f^2}{2}(\nabla\theta)^2\Bigg\rbrack \ .
 \label{eq:tpotkmem2}
 \end{equation}
In the MEM2, ${m_K^\ast}_{\rm ,MEM1}$ in Eq.~(\ref{eq:lin1}) should be replaced by the free kaon mass $m_K$ and ${m_K^\ast}_{\rm ,MEM1}^2$ in Eq.~(\ref{eq:lin4}) by $m_K^{\ast 2}$. The $\bar K$-meson coupling scheme in the MEM2 corresponds to that in Ref.~\cite{kpe95,sm96,mtv06,mtec06,gfgm07}.

\begin{figure}[h]
\begin{center}
\includegraphics[width=.6\textwidth]{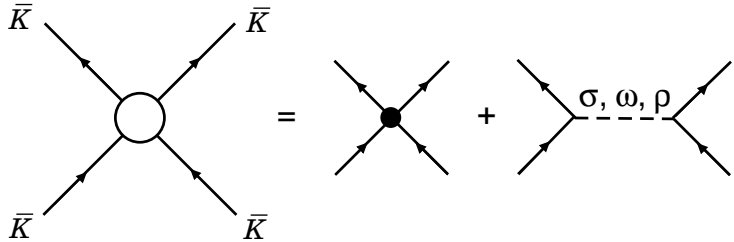}
\end{center}
\caption{Effective $\bar K-\bar K$ interaction, which consists of the contact diagram and the meson exchange diagram in the $t$ channel.}
\label{fig1}
\end{figure}

\subsection{$\bar K-\bar K$ interaction}
\label{subsec:kk}

\ \ In the chiral model, $\bar K-\bar K$ interactions are incorporated in two ways: one is originated from the nonlinear chiral Lagrangian and the other is from the meson-exchange in the $t$ channel (Fig.~\ref{fig1}). The latter is produced by the couplings between the $K^-$ and $\sigma$, $\omega$, and $\rho$ mesons through the self-consistent equations of motion, 
Eqs.~(\ref{eq:eom1})$-$(\ref{eq:eom4}). Thus the latter $\bar K-\bar K$ interaction reveals not only in the chiral model but also in the MEMs. 

To show how the $\bar K-\bar K$ interaction from the meson-exchange is generated, we specifically take up the MEM1. The equation of motion for the $K^-$ field, (\ref{eq:lin4}), includes the $\bar K-N$ attractive scalar interaction in the ``effective mass'' of the $K^-$, ${m_K^\ast}_{\rm ,MEM1}(r)$, and the $\bar K-N$ attractive vector interaction in the term $X_0(r)$ for $X_0(r)>0$. The ${m_K^\ast}_{\rm ,MEM1}(r)$ depends upon the scalar mean-field $\sigma(r)$, and $X_0(r)$ depends upon the vector mean fields $\omega_0(r)$ and $R_0(r)$. 
These mean fields are coupled to the $K^-$ field through the
kaon-coupling terms on the r.h.s. of Eqs.~(\ref{eq:lin1}),
(\ref{eq:lin2}) and (\ref{eq:lin3}). 
By eliminating the meson mean fields in ${m_K^\ast}_{\rm ,MEM1}(r)$
and $X_0(r)$ 
with the use of Eqs.~(\ref{eq:lin1}) and (\ref{eq:lin2}), (\ref{eq:lin3}), respectively, one obtains
\begin{subequations}\label{eq:elim}
\begin{eqnarray}
{m_K^\ast}_{\rm ,MEM1}(r)&=&m_K+\frac{g_{\sigma
 K}}{m_\sigma^2}\Bigg\lbrack \frac{dU}{d\sigma}-g_{\sigma
 N}(\rho_n^s+\rho_p^s)
-g_{\sigma K}{m_K^\ast}_{\rm ,MEM1} f^2\theta^2\Bigg\rbrack \ ,  \label{eq:elims} \\
X_0(r)&=&\frac{g_{\omega N} g_{\omega K}}{m_\omega^2}(\rho_p+\rho_n)+\frac{g_{\rho N} g_{\rho K}}{m_\rho^2}(\rho_p-\rho_n) \cr
&-&\Bigg(\frac{g_{\omega K}^2}{m_\omega^2}+\frac{g_{\rho K}^2}{m_\rho^2}\Bigg)(\widetilde\omega_{K^-}+X_0)f^2\theta^2 \ , \label{eq:elimv}
\end{eqnarray}
\end{subequations}
where we neglect the kinetic terms for the meson mean-fields.
The third term in the bracket on the r.~h.~s. of
Eq.~(\ref{eq:elims}), stemming from the coupling between the $\sigma$ and $K^-$ mesons, corresponds to the $\bar K-\bar K$ interaction
mediated by the $\sigma$-meson exchange, and it contributes to a
decrease of ${m_K^\ast}_{\rm ,MEM1}(r)$ as an attractive interaction. The third term on the r.~h.~s. of Eq.~(\ref{eq:elimv}), stemming from the coupling between the $\omega$, $\rho$ and $K^-$ mesons, corresponds to the $\bar K-\bar K$ interaction mediated by the $\omega$ and $\rho$-mesons exchange, 
and it contributes to a decrease of $X_0(r)$ as a repulsive interaction. 
By substituting Eqs.~(\ref{eq:elims}) and (\ref{eq:elimv}) into the
equation of motion for $\theta$, (\ref{eq:lin4}), 
one can see that these two terms render another kind of
nonlinear $\bar K-\bar K$ interaction terms other than those coming from the nonlinear representation of the $K^-$ field. 

In contrast to the chiral model adopted in this paper and MEMs, 
 the previous results on kaon condensation 
in neutron-star matter are based on the nonlinear chiral Lagrangian, where the $K^-$ mesons 
do not couple with the scalar or vector mesons dynamically so that the $\bar K-\bar K$ interaction originates solely from the nonlinear representation of the $K^-$ field. In that case, the $\bar K-N$ attractive scalar interaction is given in terms of the $KN$ $\sigma$ term, $\Sigma_{KN} $, through $m_K^{\ast 2}(r)=m_K^2-\Sigma_{KN}(\rho_p^s+\rho_n^s)/f^2$, which does not include the $\bar K-\bar K$ attractive term corresponding to the third term in the bracket on the r.~h.~s. of Eq.~(\ref{eq:elims}). In addition the 
$\bar K-N$ attractive vector interaction (the 
Tomozawa-Weinberg term) is given by $X_0=(\rho_p+\rho_n/2)/(2f^2)$, which does not include the $\bar K-\bar K$ repulsive term corresponding to the third term on the r.~h.~s. of Eq.~(\ref{eq:elimv}) and increases monotonically with baryon density 
as kaon condensation develops\cite{kn86,mtt93,lbm95,fmmt96,tpl94}.
For a quantitative discussion, see Sec.~\ref{subsubsec:3-2-3}. 

\section{Numerical results and Discussion}
\label{sec:results}

\subsection{Case for $|S|$=0, 1, 2}
\label{subsec:1}

\ \ First we consider the cases of $|S|$=0, 1, 2, where the nonlinear $\bar K-\bar K$ terms are absent or they are expected to have a minor effect on the properties of the MKN. 

\begin{figure}[h]
\begin{center}
\includegraphics[height=.3\textheight]{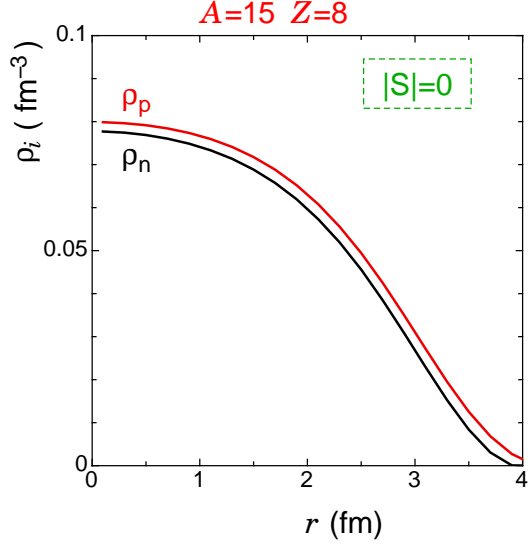}
\end{center}
\caption{The density distributions of the protons and neutrons for the nucleus $^{15}_{\ 8}$O ($A$=15, $Z$=8, and $|S|$=0).}
\label{fig2}
\end{figure}

\subsubsection{Density profiles}
\label{subsubsec:density}

\ \ We take a reference nucleus with no trapped $K^-$ meson to $^{15}_{\ 8}$O ($A$=15, $Z$=8). The density profile for $^{15}_{\ 8}$O is shown in Fig.~\ref{fig2}. The central baryon density $\rho_{\rm B}^{(0)}$ [=$\rho_p(r=0)+\rho_n(r=0)$] is 1.03$\rho_0$, and the densities of the protons and neutrons are almost equally distributed. The binding energy per nucleon is evaluated as 6.93 MeV/A, which is less than but close to the empirical value, 7.46 MeV/A\cite{awt03}.

\begin{figure}[!]
\noindent
\begin{minipage}[l]{0.50\textwidth}
\begin{center}
\includegraphics[height=.3\textheight]{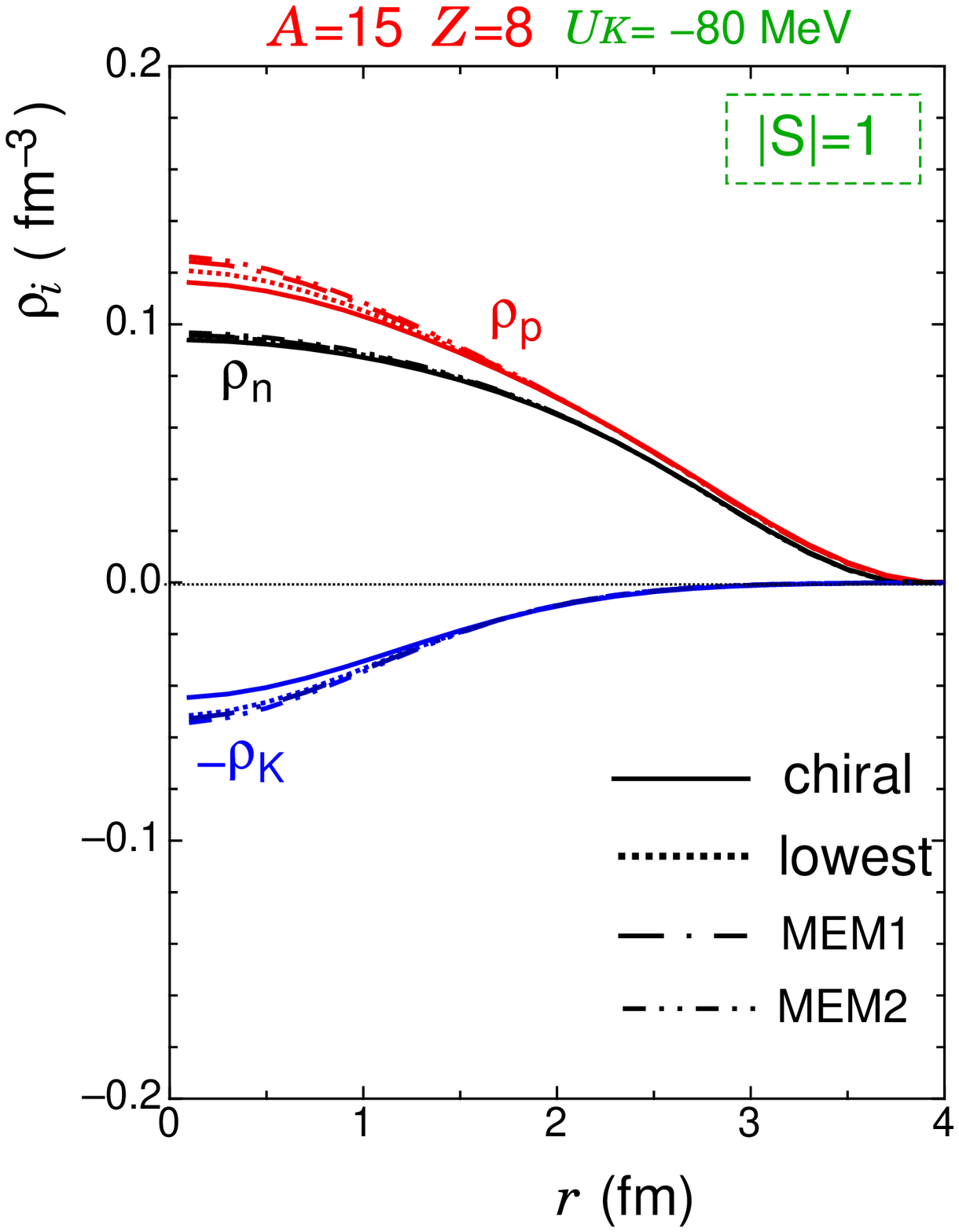}
\end{center}
\end{minipage}~
\begin{minipage}[r]{0.50\textwidth}
\begin{center}
\includegraphics[height=.3\textheight]{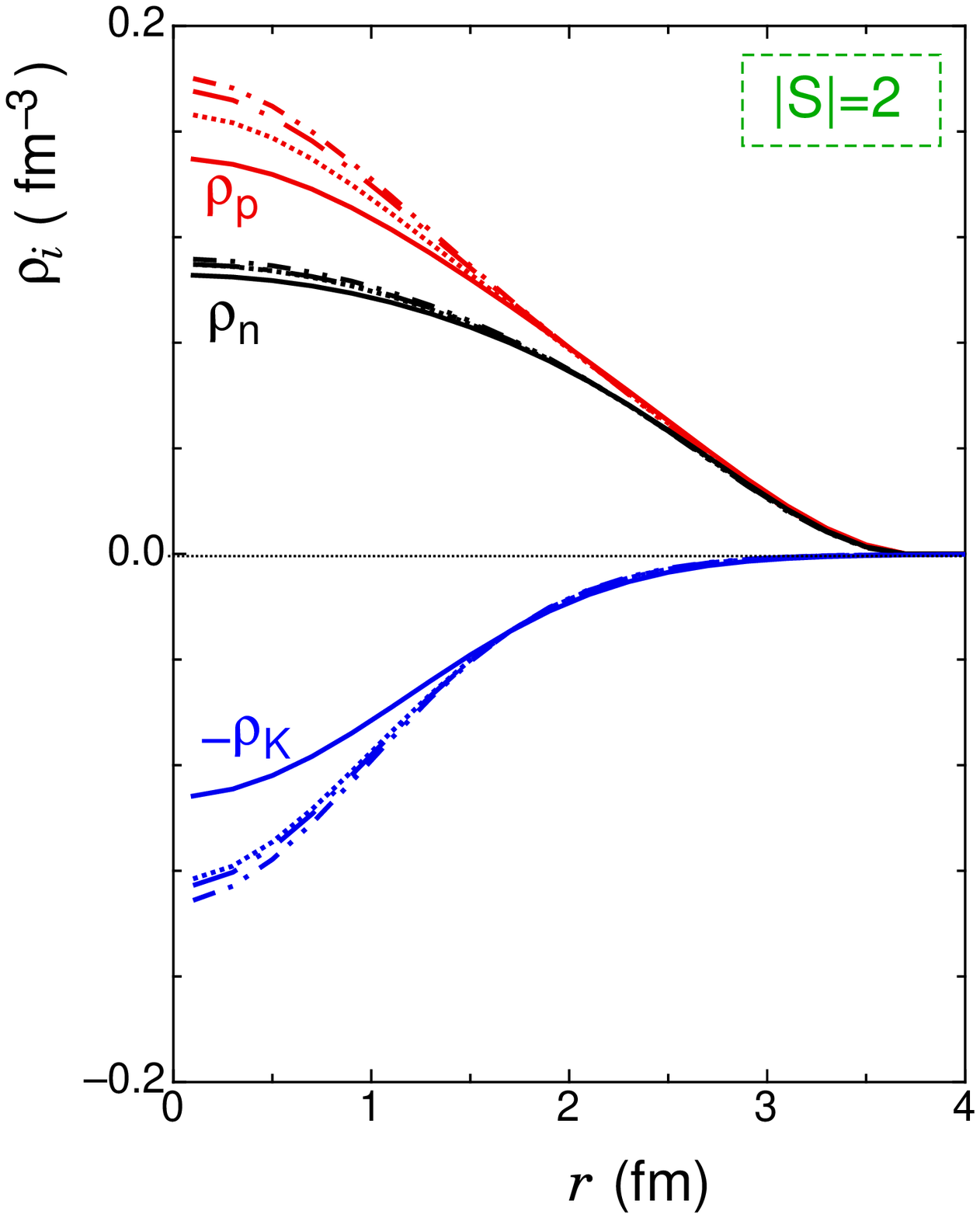}
\end{center}
\end{minipage}
\caption{The density distributions of the protons, neutrons, and the distribution of the strangeness density [=$-\rho_{K^-} (r)$] are shown for the MKN with $A$=15, $Z$=8, and $|S|$=1, 2 in the case of $U_K$=$-$80 MeV. The solid lines are for the chiral model, dashed-dotted lines for MEM1, dashed-two-dotted lines for MEM2. The results from the lowest-order approximation of the $\Omega$ with respect to the $K^-$ field are also shown by the dotted lines.  }
\label{fig3}
\end{figure}

\begin{figure}[!]
\noindent
\begin{minipage}[l]{0.50\textwidth}
\begin{center}
\includegraphics[height=.3\textheight]{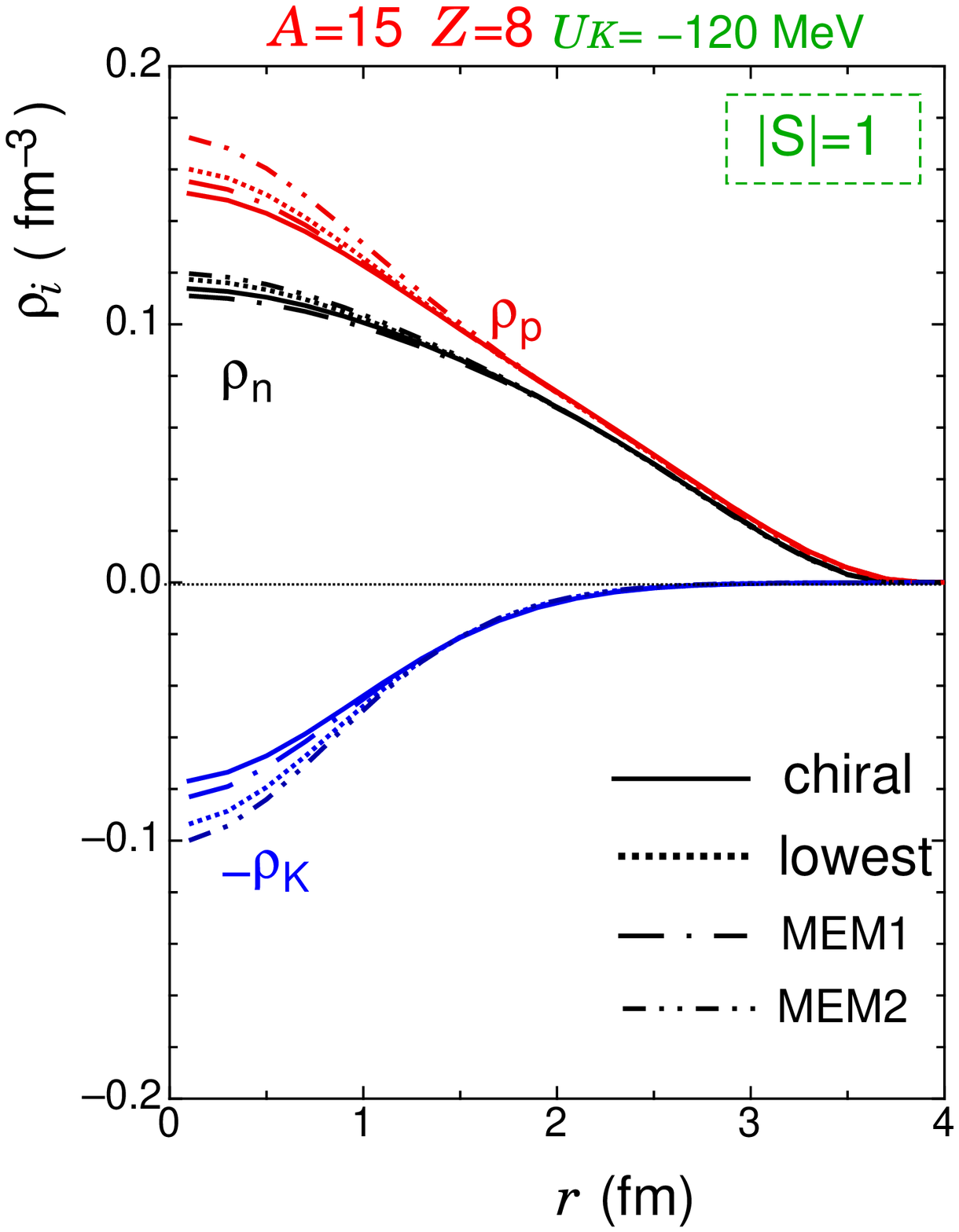}
\end{center}
\end{minipage}~
\begin{minipage}[r]{0.50\textwidth}
\begin{center}
\includegraphics[height=.3\textheight]{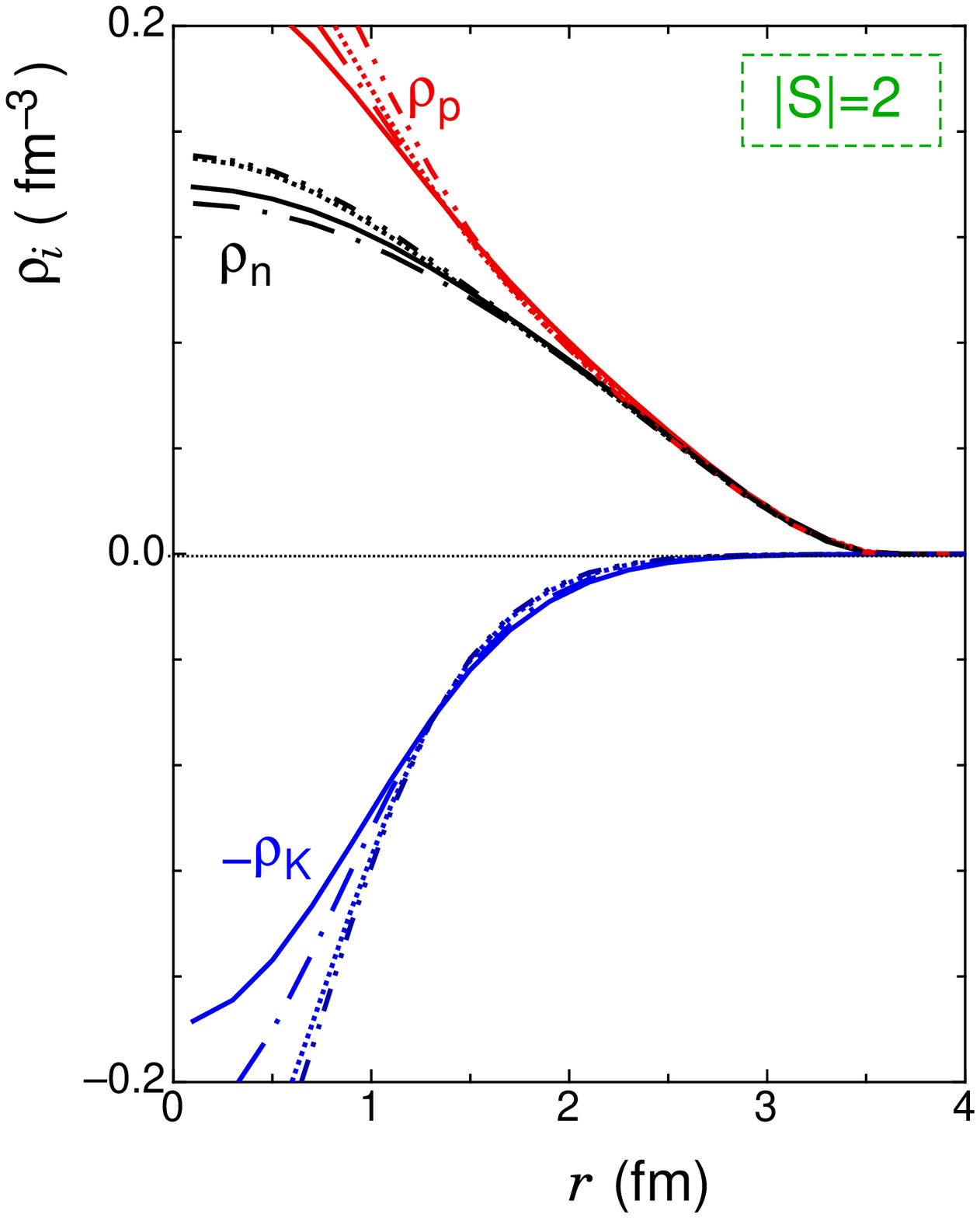}
\end{center}
\end{minipage}
\caption{The same as Fig.~\ref{fig3} but for $U_K$=$-$120 MeV. See the text for details. }
\label{fig4}
\end{figure}

 In Figs.~\ref{fig3} and \ref{fig4}, we show the density distributions of the protons ($\rho_p(r)$), neutrons ($\rho_n(r)$), and the distribution of the strangeness density [=$-\rho_{K^-} (r)$] for the MKN with $A$=15, $Z$=8, and $|S|$=1, 2 in the case of the $K^-$ optical potential depth $U_K$=$-$80 MeV and $-$120 MeV, respectively. 
The solid lines are for the chiral model, the dashed-dotted lines for MEM1, and the dashed-two-dotted lines for MEM2. 
For all the cases of the chiral and meson-exchange models, MEM1 and MEM2, the density distribution of the $K^-$ [$\rho_{K^-} (r)$] has a peak at the center of the nucleus and decreases monotonically with the radial distance $r$. Both the protons and neutrons are attracted to the $K^-$. Since the $\bar K N$ attractive interaction is stronger for the isospin $I$ = 0 than for $I$ = 1, the protons are more strongly attracted to the center than the neutrons. 

\begin{figure}[!]
\begin{minipage}[l]{0.50\textwidth}
\begin{center}
\includegraphics[height=.3\textheight]{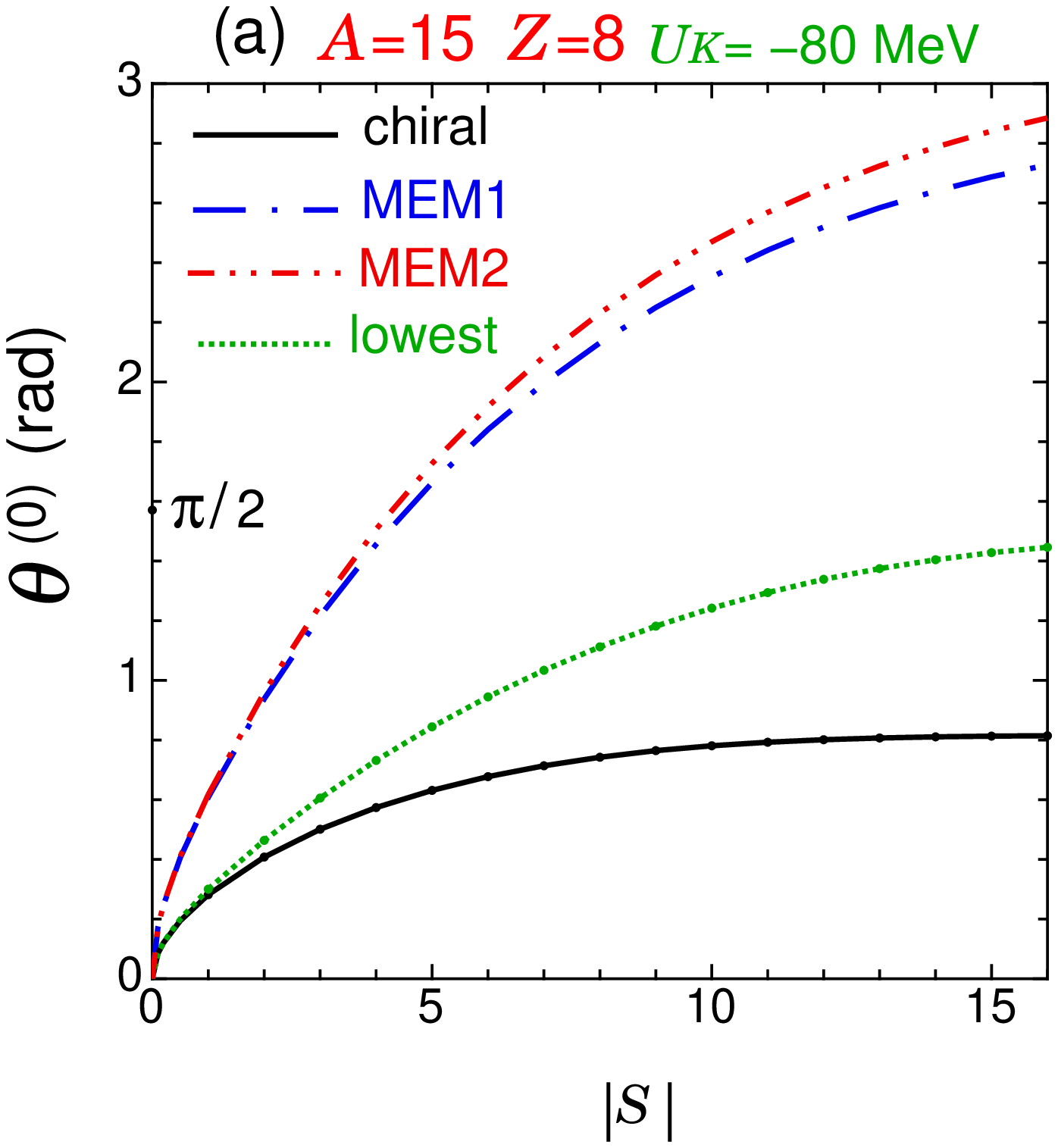}\end{center}
\end{minipage}~
\begin{minipage}[r]{0.50\textwidth}
\begin{center}
\includegraphics[height=.3\textheight]{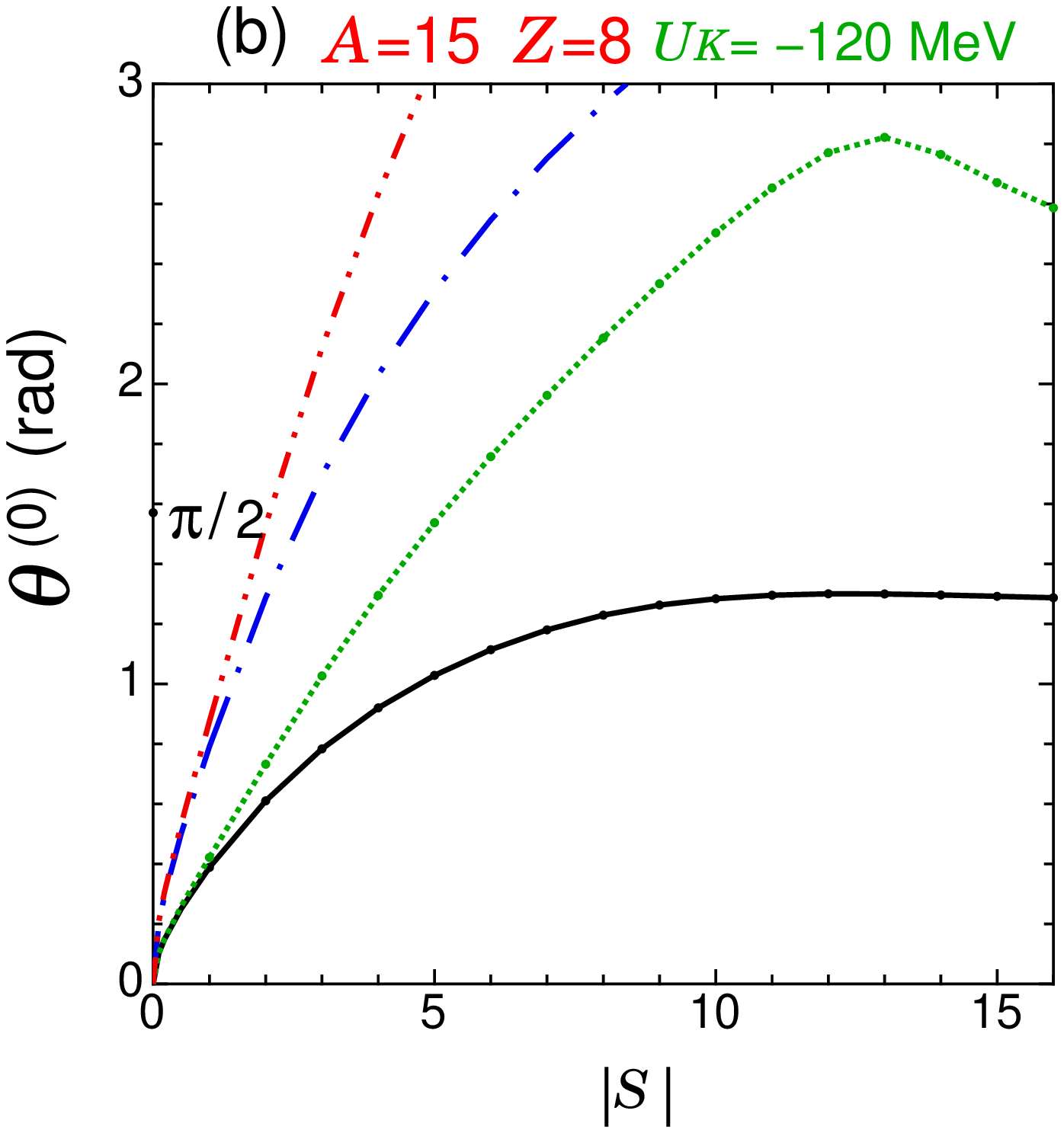}
\end{center}
\end{minipage}
\caption{(a) The chiral angle at the center of the MKN, $\theta^{(0)} [=\theta(r=0)$], as functions of $|S|$ for $U_K$=$-$80 MeV. The $^{15}_{\ 8}$O ($A$=15, $Z$=8) is taken as a reference nucleus. The meaning of the curves is the same as in Fig.~\ref{fig3}. (b) The same as (a) but for $U_K$=$-$120 MeV.
\protect\label{fig5}}
\end{figure}

To discuss roles of the nonlinear $\bar K-\bar K$ terms in the chiral model, we also present the results with an approximation where only the expansion, $\sin\theta\rightarrow \theta$, $\cos\theta\rightarrow 1-\theta^2/2$, is done for the $\Omega$ in the chiral model without any additional terms (i) and (ii). We call it the ``lowest-order approximation'' (abbreviated as ``lowest'' in the Figures). For a small chiral angle ($\theta (r)\lesssim 0.5$ (rad) ), the difference between the chiral model and the lowest-order approximation is expected to be small. We show, in Fig.~\ref{fig5}, the results of the chiral angle at the center of the MKN, $\theta^{0} [=\theta(r=0)$], as functions of $|S|$ for (a) $U_K$=$-$80 MeV and (b) $U_K$=$-$120 MeV. The solid lines are for the chiral model, and the dotted lines are for the lowest-order approximation. 
The $\theta^{(0)}$ is small and is not much different between the chiral model and the lowest-order approximation for small values of $|S|$ (=1,2) for both $U_K$=$-$80 MeV and $-$ 120 MeV. Therefore, there is no significant difference in the density distributions $\rho_i(r)$ ($i=p, n, K^-$) between the chiral model and the lowest-order approximation for small values of $|S|$, as seen in Figs.~\ref{fig3} and \ref{fig4} (the solid and the dotted lines). 

The difference between the results for the MEMs and those for the lowest-order approximation comes from the additional two terms, both of which are of $O(\bar K^2)$ [the $\bar K \bar K VV$ term (i) and the $\bar K \bar K \sigma\sigma$ term (ii) for the MEM1 and the $\bar K \bar K VV$ term (i) for the MEM2] . 
In fact, there is a definite difference of the $\theta^{(0)}$ between the lowest-order approximation or the chiral model and the MEMs even for  $|S|$ =1,2, as seen in Fig.~\ref{fig5}. 
Qualitatively, due to the attractive contribution from (i), the $K^-$ mesons, protons, and neutrons are attracted more to the center of the MKN for  the MEM2 than for the lowest-order approximation (the dashed-two-dotted lines and dotted lines in Figs.~\ref{fig3} and \ref{fig4}), while this tendency is hindered for the MEM1 due to the repulsive contribution from (ii)  (the dashed-dotted lines and dotted lines in Figs.~\ref{fig3} and \ref{fig4}). 
Nevertheless these two terms (i) and (ii) do not produce a large difference of the density distributions $\rho_i(r)$ ($i=p, n, K^-$) between the lowest-order approximation (or the chiral model) and the MEMs. 

\subsection{Case for $|S|\geq$ 3}
\label{subsec:3-2}

\ \ Next we consider the effect of the nonlinear $\bar K-\bar K$ terms  on the properties of the MKN in the case of $|S|\geq$ 3. 

\begin{figure}[ph]
\noindent\begin{minipage}[l]{0.50\textwidth}
\begin{center}
\includegraphics[height=.3\textheight]{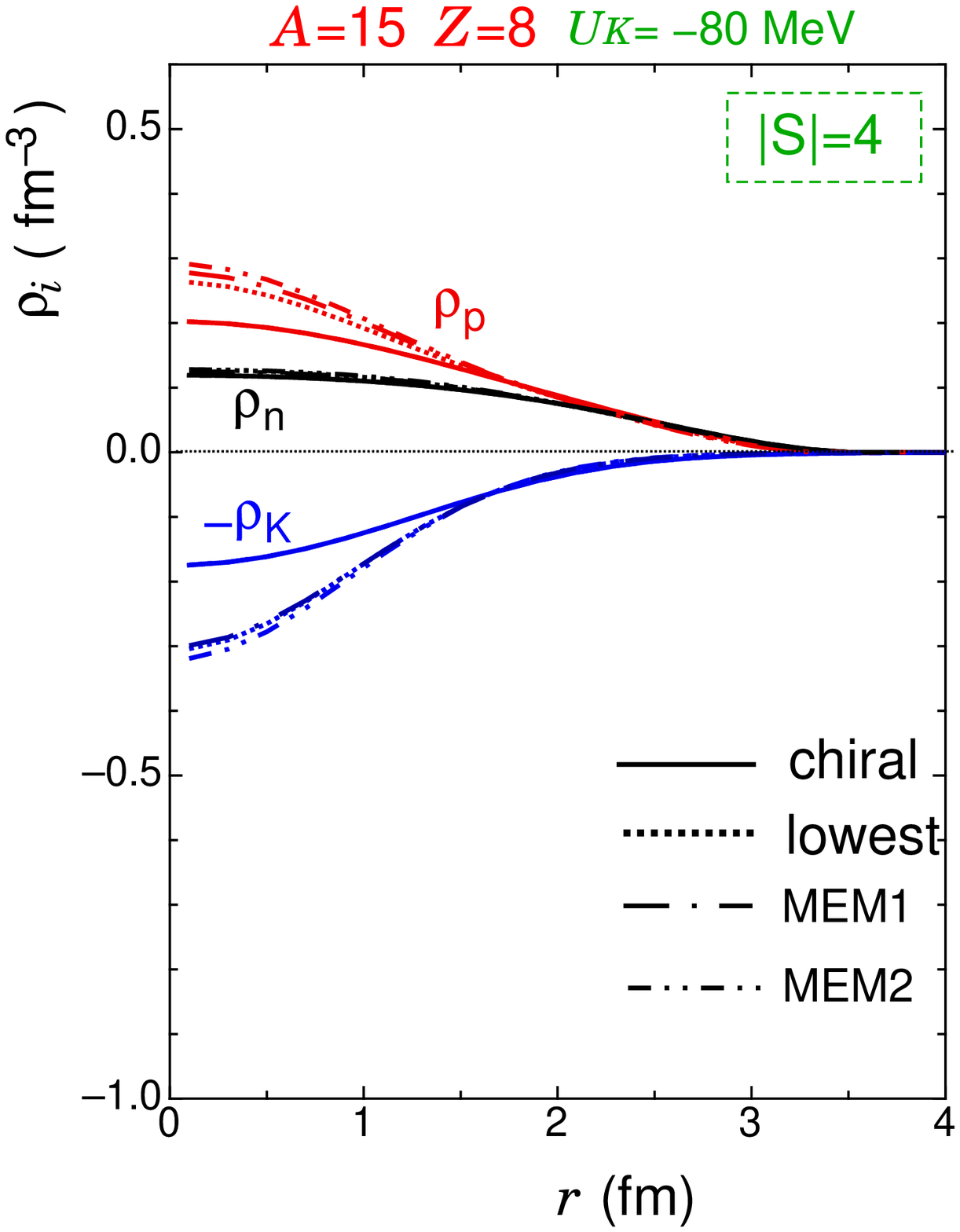}
\end{center}
\end{minipage}~
\begin{minipage}[r]{0.50\textwidth}
\begin{center}
\includegraphics[height=.3\textheight]{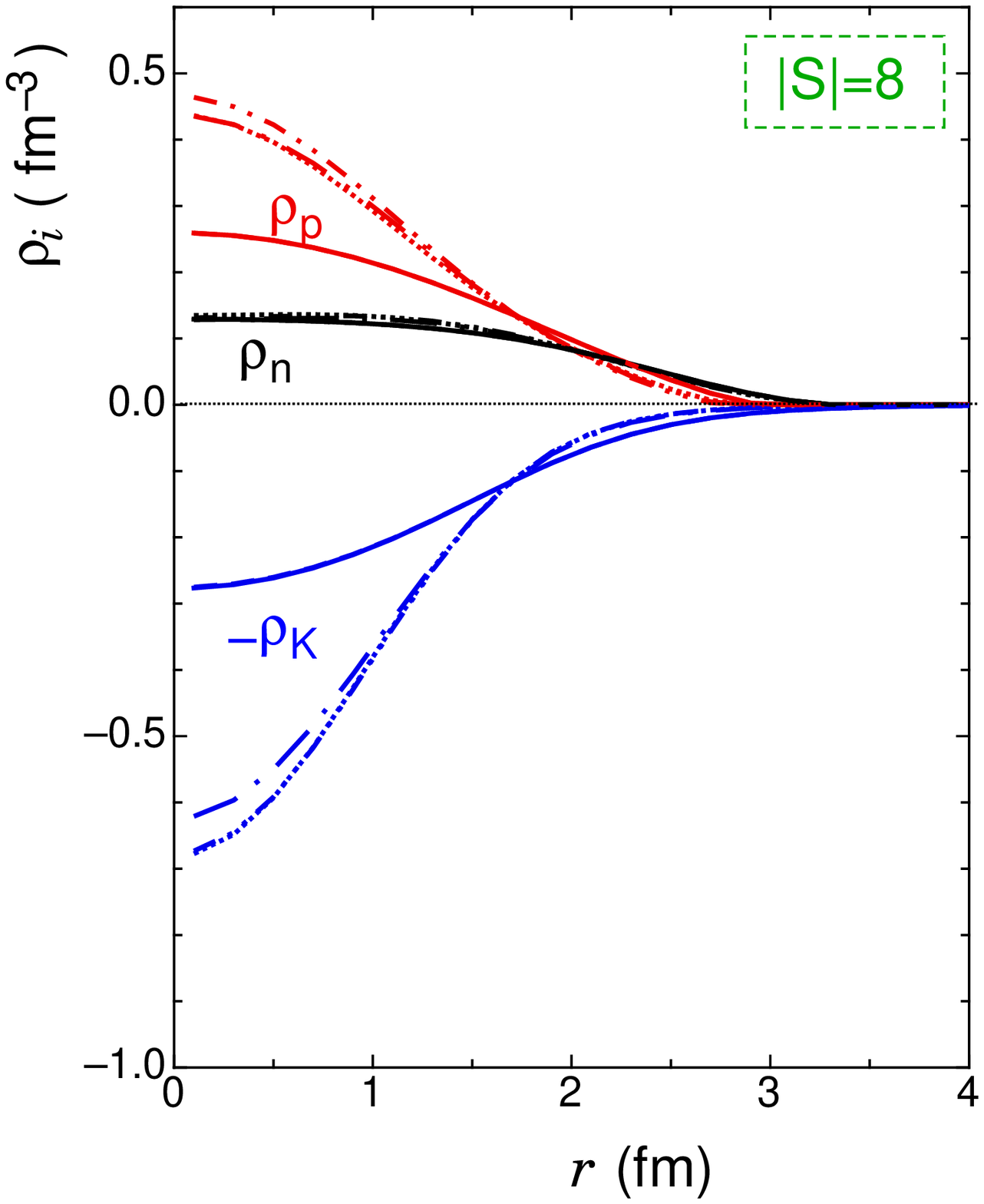}
\end{center}
\end{minipage}
\vspace{0.3cm}

\noindent\begin{minipage}[l]{0.50\textwidth}
\begin{center}
\includegraphics[height=.3\textheight]{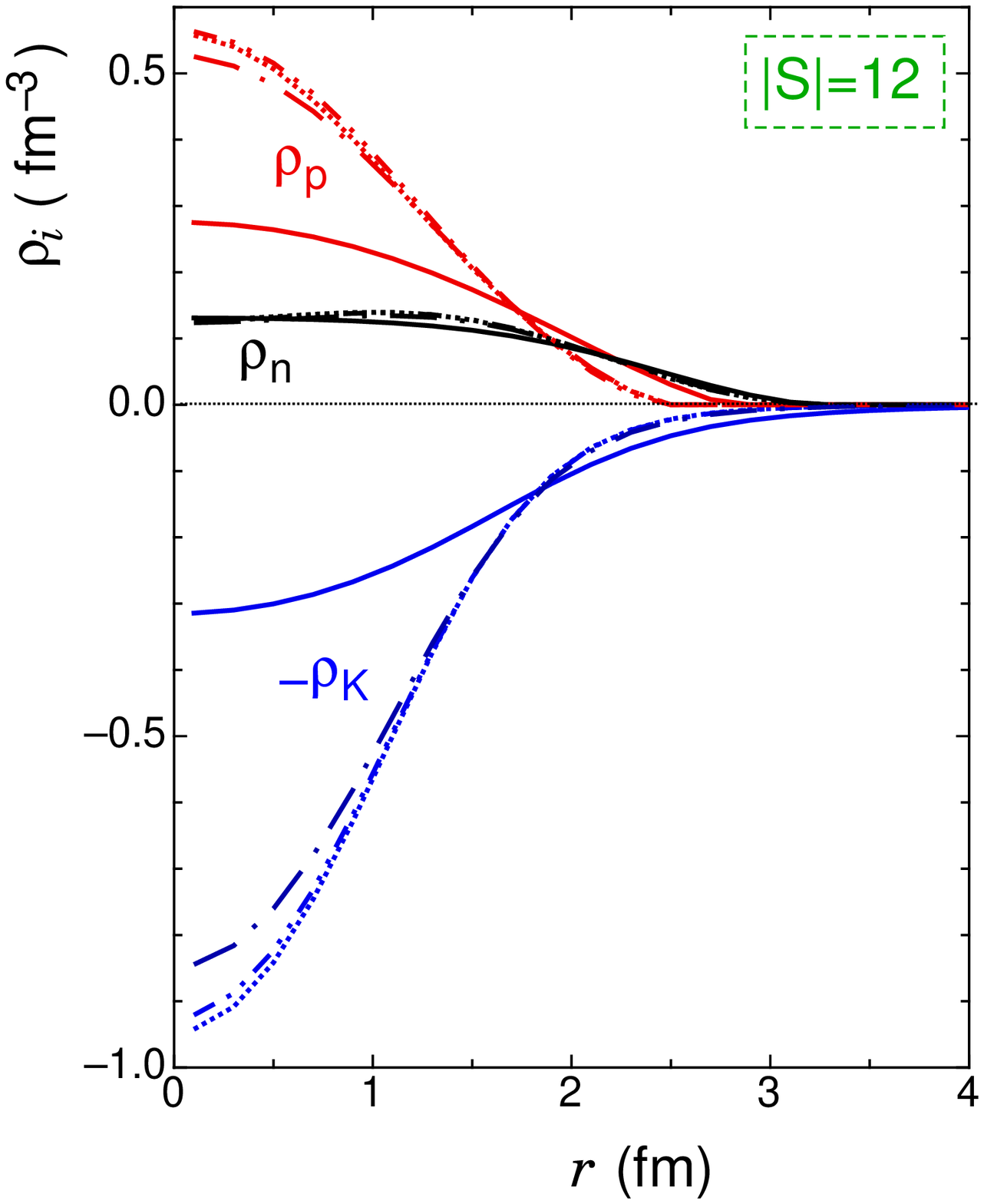}
\end{center}
\end{minipage}~
\begin{minipage}[r]{0.50\textwidth}
\begin{center}
\includegraphics[height=.3\textheight]{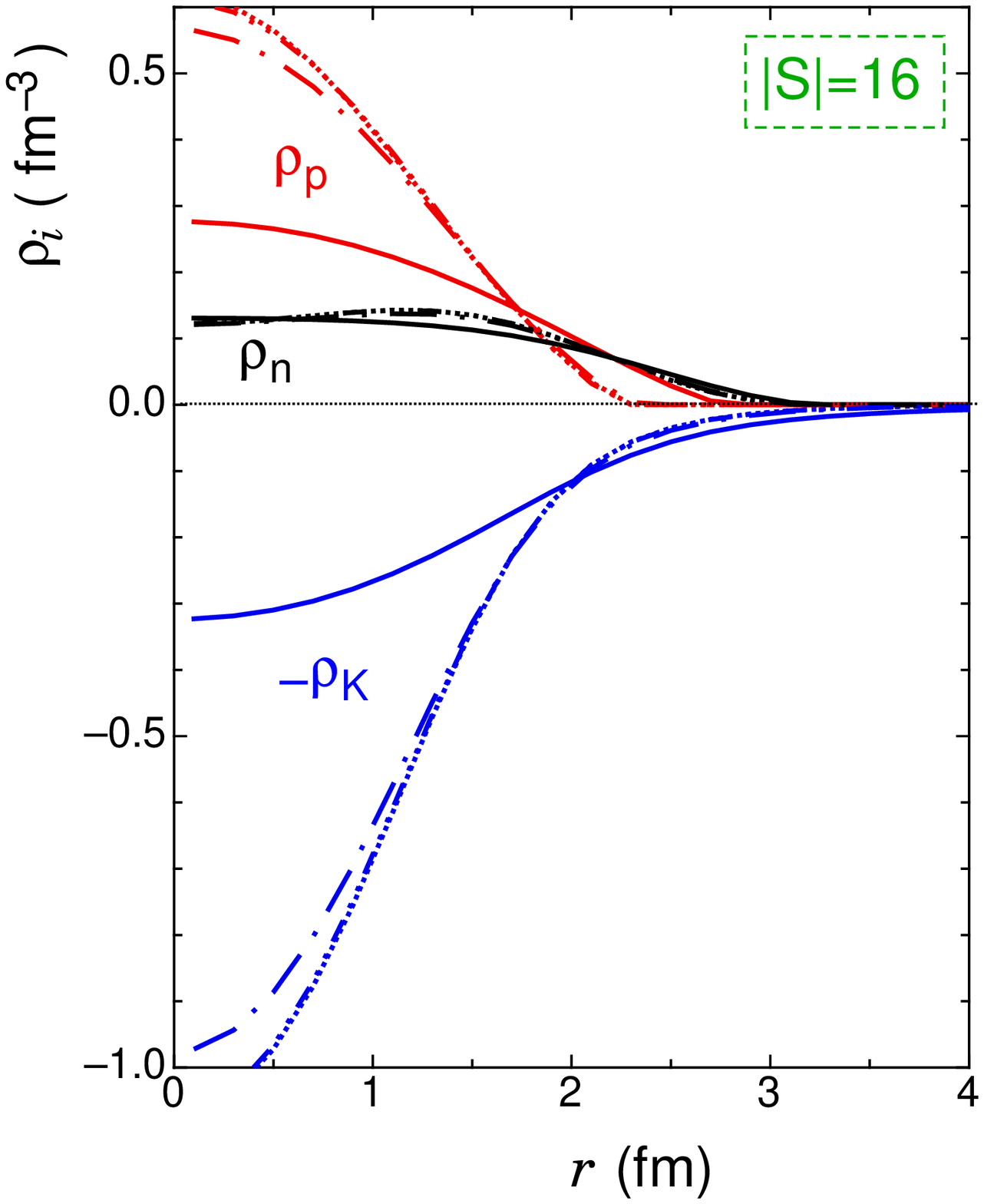}
\end{center}
\end{minipage}
\vspace*{8pt}
\caption{The same as Fig.~\ref{fig3} but for $|S|$= 4, 8, 12, 16.
\protect\label{fig6}}
\end{figure}

\begin{figure}[ph]
\noindent\begin{minipage}[l]{0.50\textwidth}
\begin{center}
\includegraphics[height=.3\textheight]{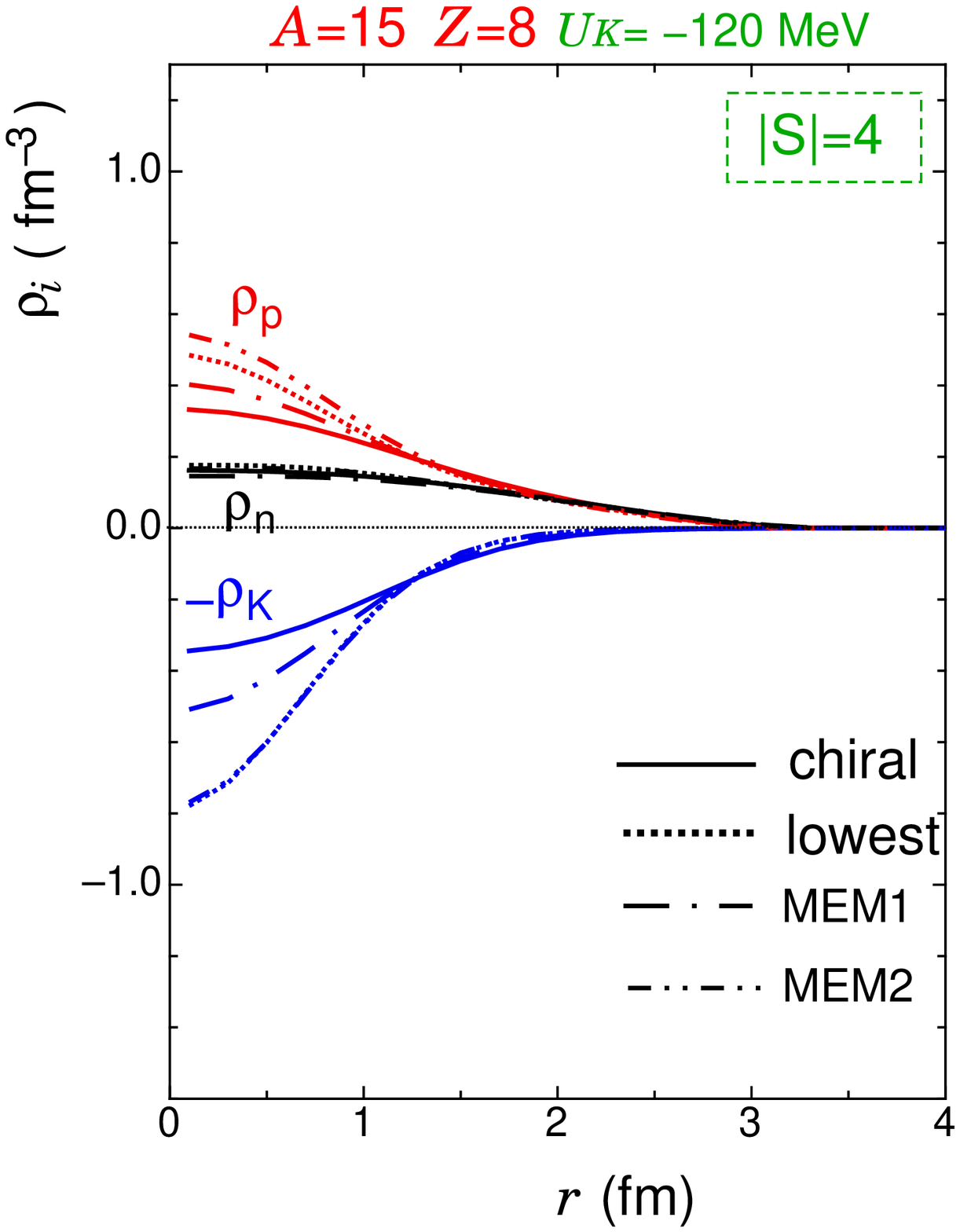}
\end{center}
\end{minipage}~
\begin{minipage}[r]{0.50\textwidth}
\begin{center}
\includegraphics[height=.3\textheight]{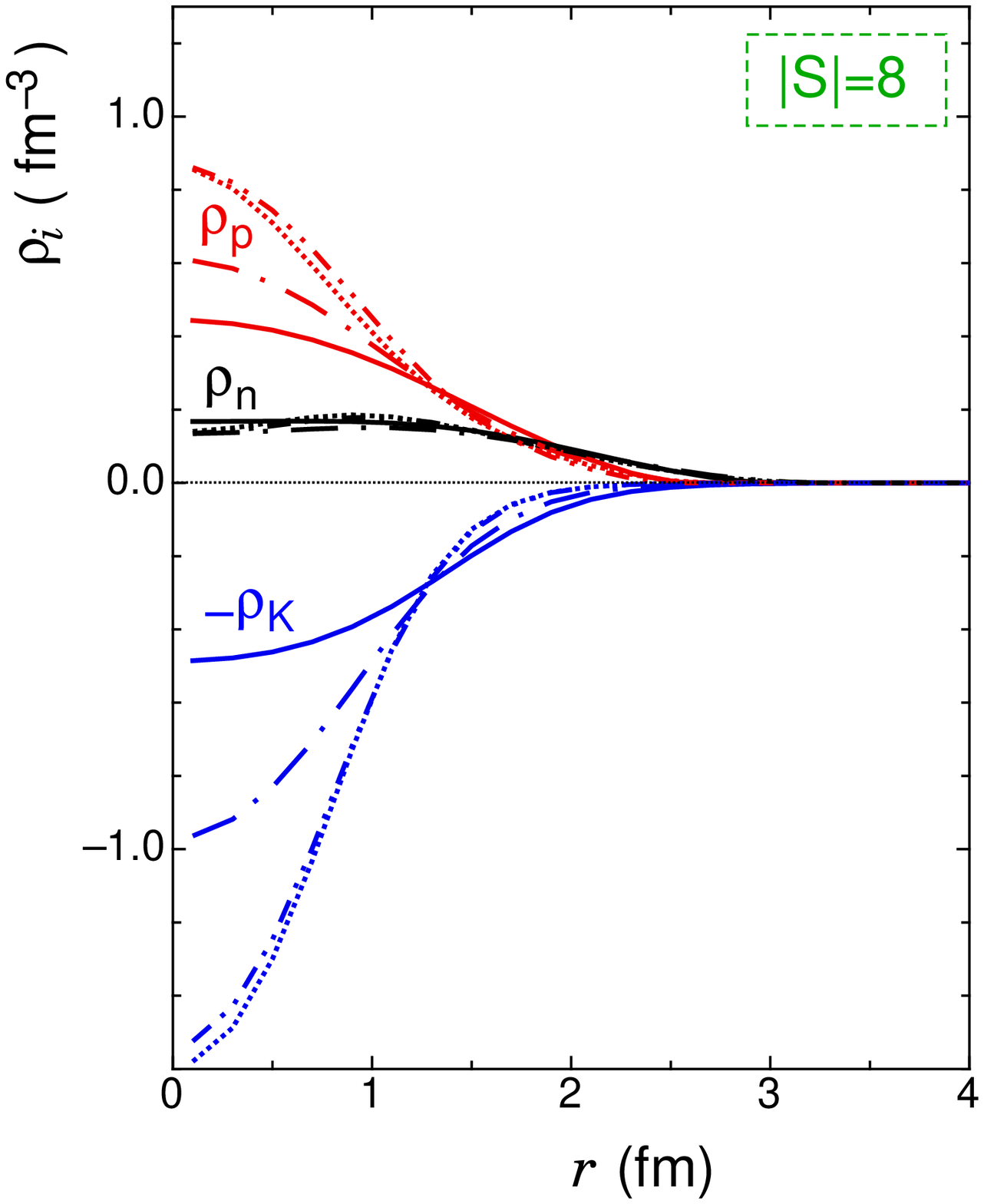}
\end{center}
\end{minipage}
\vspace{0.3cm}

\noindent\begin{minipage}[l]{0.50\textwidth}
\begin{center}
\includegraphics[height=.3\textheight]{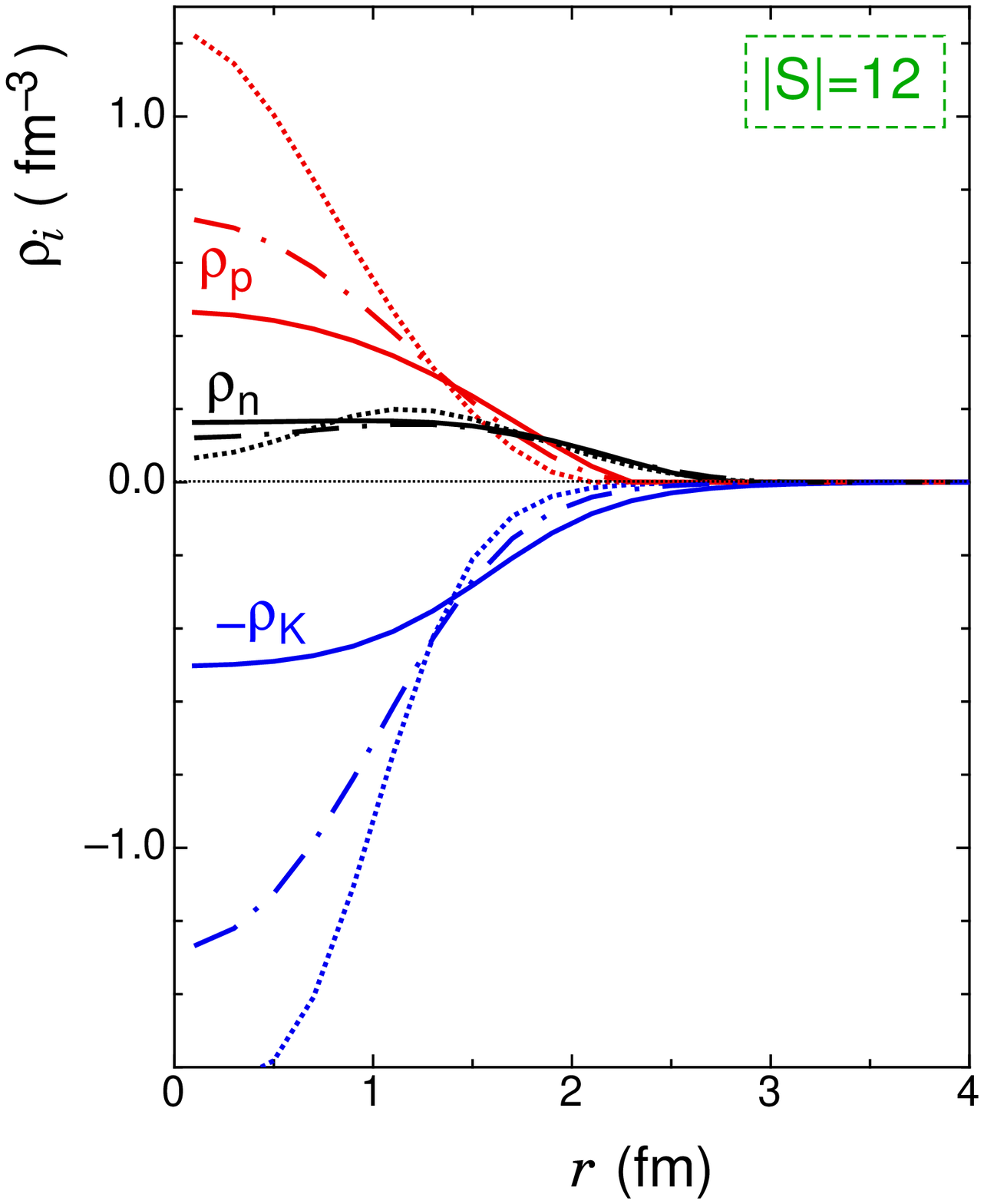}
\end{center}
\end{minipage}~
\begin{minipage}[r]{0.50\textwidth}
\begin{center}
\includegraphics[height=.3\textheight]{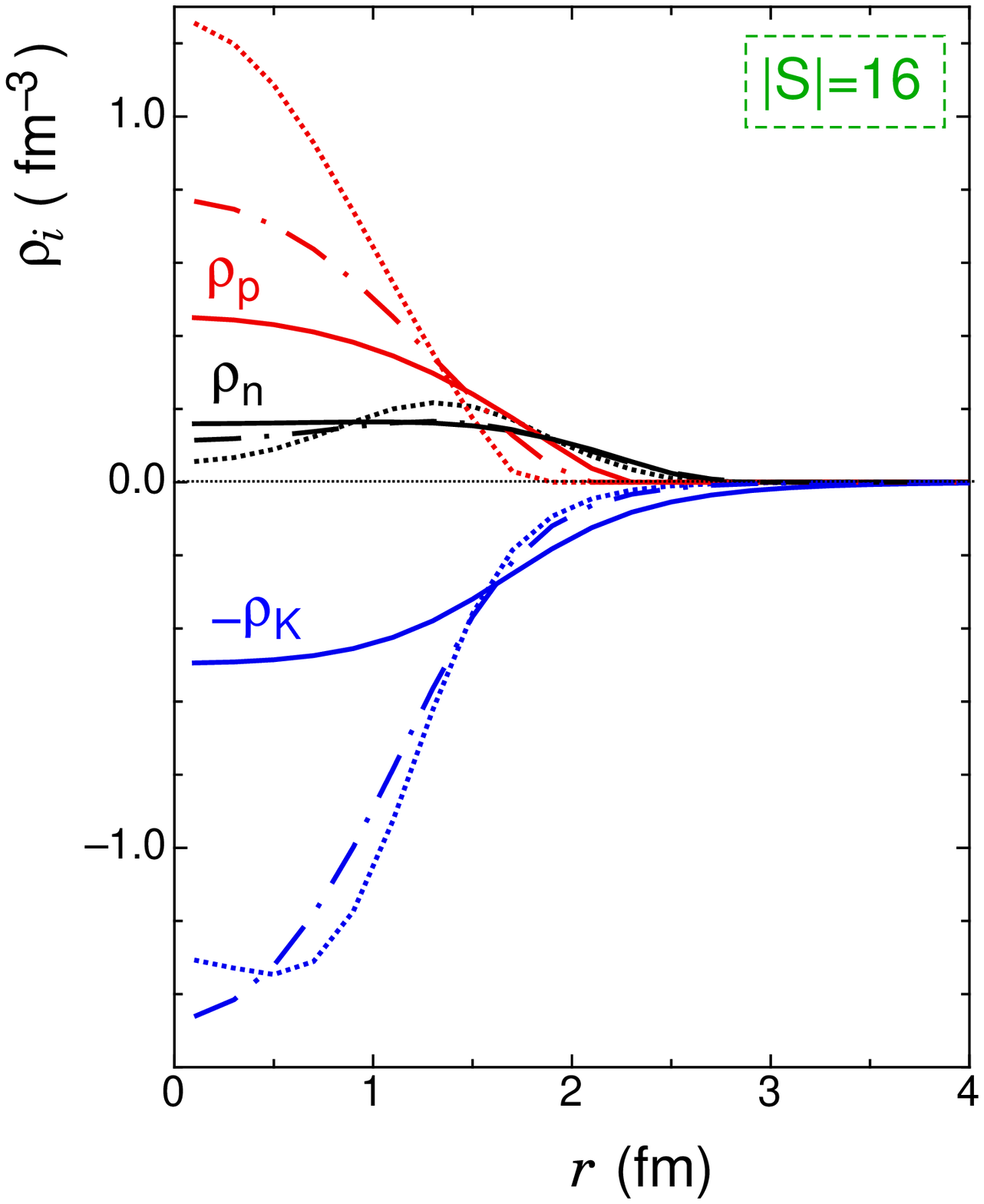}
\end{center}
\end{minipage}
\vspace*{8pt}
\caption{The same as Fig.~\ref{fig6} but for $U_K$=$-$120 MeV. Note that the stable solutions for the MEM2 are not obtained for $|S|\geq$12. 
\protect\label{fig7}}
\end{figure}

\subsubsection{Density profiles}
\label{subsubsec:3-2-1}

\ \ In Figs.~\ref{fig6} and \ref{fig7}, we show the density distributions of the protons ($\rho_p(r)$), neutrons ($\rho_n(r)$), and the distribution of the strangeness density [=$-\rho_{K^-} (r)$] for the MKN with $A$=15, $Z$=8, and $|S|$=4, 8, 12, 16 in the case of $U_K$=$-$80 MeV and $-$120 MeV, respectively. Note that the stable solutions for the MEM2 are not obtained for $|S|\geq$12 in the case of $U_K$=$-$120 MeV. 

In the case of the chiral model (the solid lines), the $K^-$ distribution is pushed outward and tends to be uniform with the increase in $|S|$, in comparison with the cases of the MEM1 (the dashed-dotted lines), MEM2 (the dashed-two-dotted lines), and the lowest-order approximation (the dotted lines).
This is because the nonlinear $\bar K-\bar K$ repulsive interaction gets strong as the number of $K^-$ mesons increases,  leading to repel the $K^-$ mesons with each other. In the case of the chiral model, one can also see that the nucleon density distributions [$\rho_p (r)$ and $\rho_n (r)$] are saturated for $|S|\gtrsim$ 8,  keeping from the increase of both the nucleon Fermi energies and short-range repulsive interaction between nucleons. In particular, the neutron distribution becomes almost uniform over the large part of the MKN. 
The results on the density distributions of the protons, neutrons, and $K^-$ mesons for $U_K$=$-$120 MeV are qualitatively the same as the above results for $U_K$=$-$80 MeV. [See also Ref.~\cite{mmt07}.]
It should be pointed out that a ``neutron skin'' structure appears on the tails of the nucleon density profiles for a large $|S|$ ($\gtrsim$ 6) for all the adopted models (Figs.~\ref{fig6} and \ref{fig7}). 
In the chiral model, the proton distribution is relatively smeared outward in accordance with the $K^-$ distribution in comparison with  other models due to the nonlinear $\bar K-\bar K$ repulsion, so that the difference between the root mean square (RMS) radii of the neutron and proton becomes small. Therefore,
the neutron skin structure can be seen more distinctly in the cases of the MEMs and the lowest-order approximation than the case of the chiral model. 
It is summarized that the density profiles of the proton, neutron, and  $K^-$ meson in the chiral model are quite unique as compared with those in the MEMs and in the lowest-order approximation as a result of the repulsive effects originating from the nonlinear $\bar K-\bar K$ interaction terms.

To see higher order effects from the $K^-$ field in the chiral model, we consider another approximation, where the expansion of the $\Omega$ in the chiral model with respect to $\theta$ in $O(\theta^4)$ is done such that 
\begin{equation}
\sin\theta\rightarrow \theta-\theta^3/3! \ , \quad \cos\theta\rightarrow 1-\theta^2/2+\theta^4/4!
\label{eq:quartic}
\end{equation}
 without any additional terms (i) ($\bar K \bar KVV$ term) and 
(ii) ($\bar K \bar K\sigma\sigma$ term). We call it the ``$O(\bar K^4)$ approximation''. The approximation (\ref{eq:quartic}) is valid for $\theta\lesssim$1.5. As we can see in Fig.~\ref{fig5}, the values of the $\theta^{(0)}$ are less than 1.5 rad over the relevant region of $|S|$ in the chiral model, so that the $\theta^{(0)}$ and the other quantities in the chiral model are well reproduced by the $O(\bar K^4)$ approximation. Indeed, the density profiles of the proton, neutron, and $K^-$ meson obtained in the $O(\bar K^4)$ approximation are almost identical to those in the chiral model. 
This result shows that the nonlinear $\bar K-\bar K$ interaction terms are exhausted by the terms up to the quartic terms with respect to the $K^-$ field and that the terms higher order than the quartic terms are canceled out with each other. 

\begin{figure}[!]
\begin{minipage}[l]{0.50\textwidth}
\begin{center}
\includegraphics[height=.3\textheight]{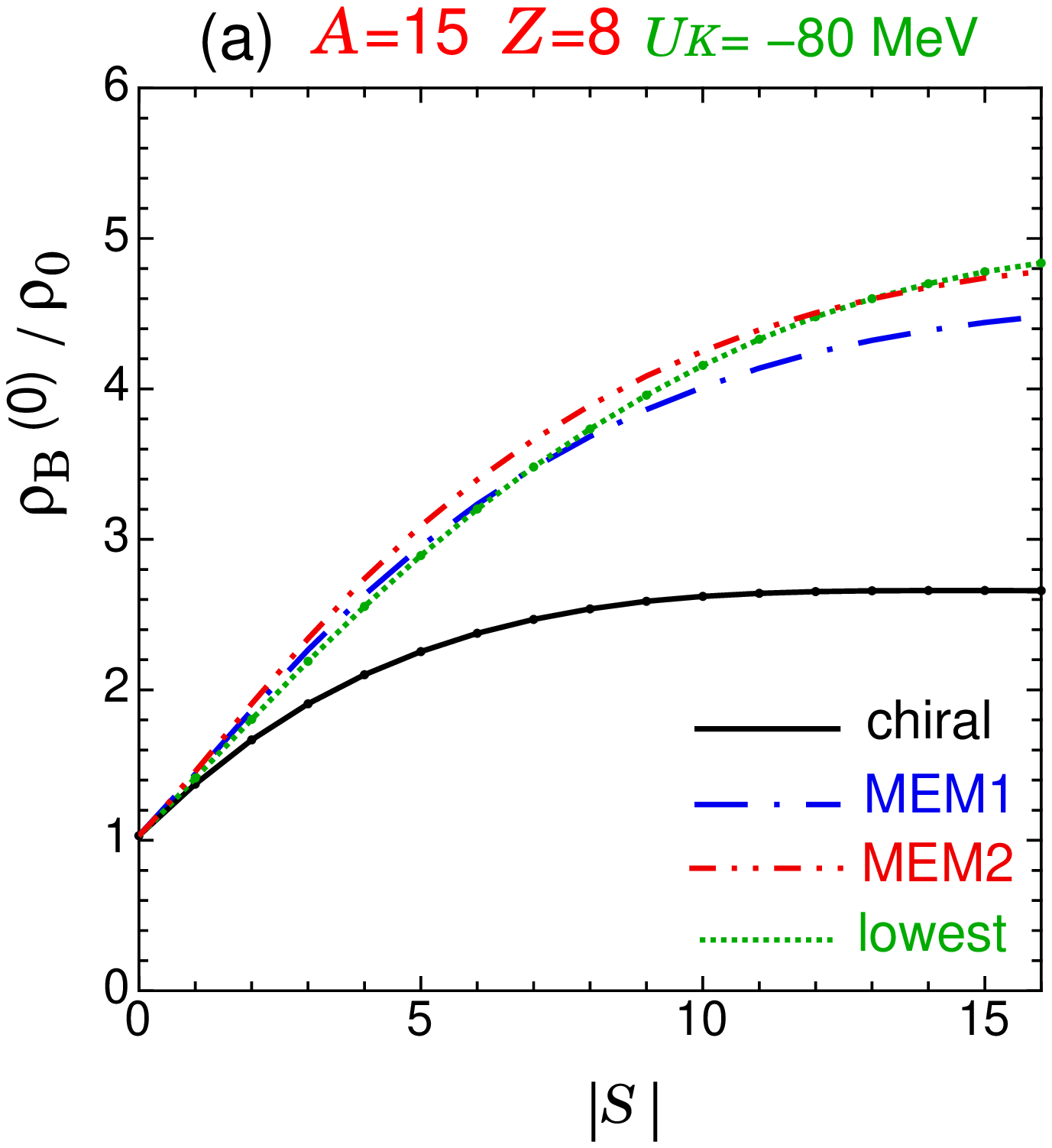}
\end{center}
\end{minipage}~
\begin{minipage}[r]{0.50\textwidth}
\begin{center}
\includegraphics[height=.3\textheight]{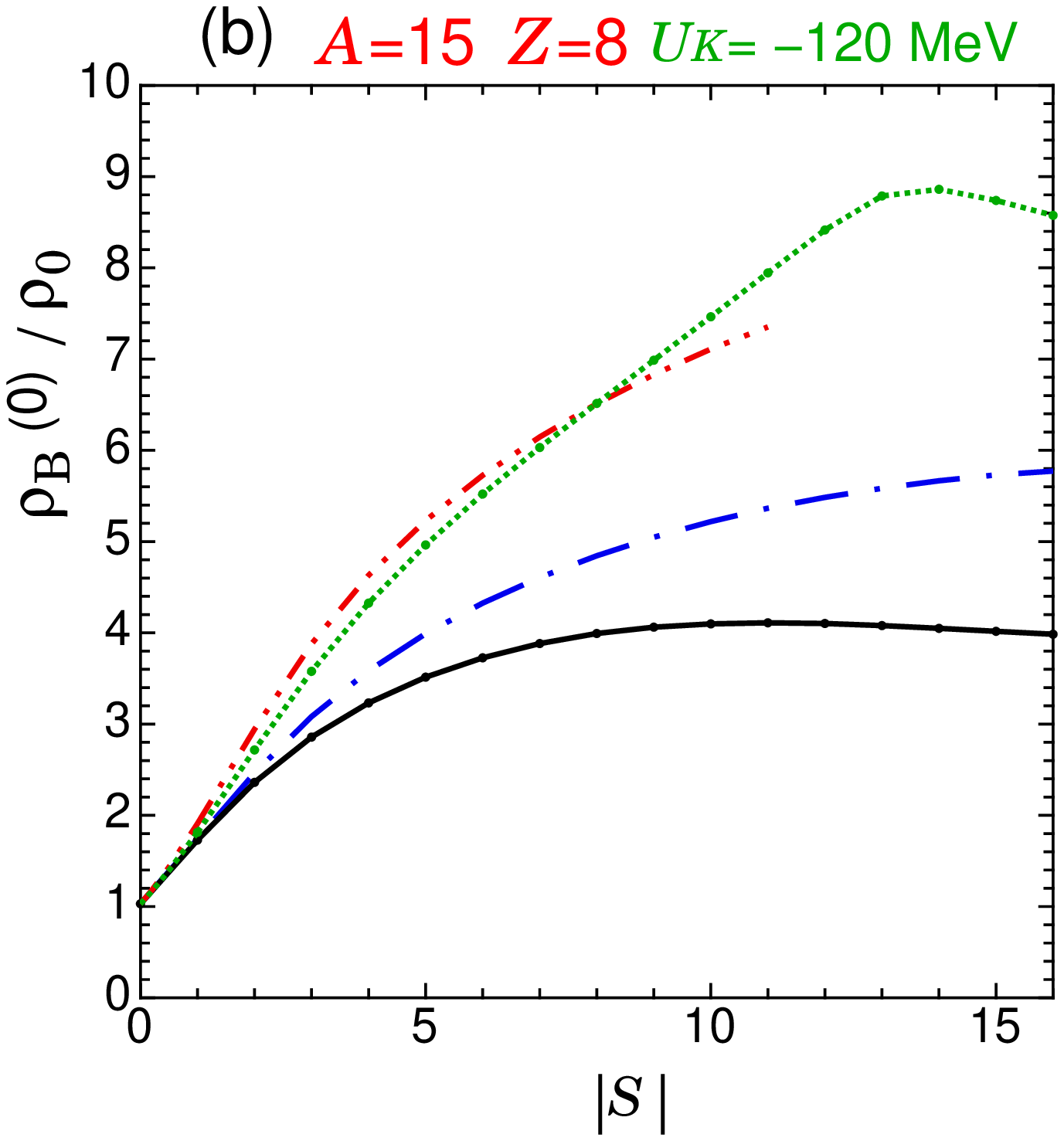}
\end{center}
\end{minipage}
\caption{(a) The baryon number density at $r=0$, $\rho_{\rm B}^{(0)}$ [=$\rho_p(r=0)+\rho_n(r=0)$], in the unit of the standard nuclear density $\rho_0$ for the MKN with $A$=15, $Z$=8 as functions of $|S|$ in the case of 
$U_K$=$-$80 MeV. The meaning of the curves is the same as in Fig.~\ref{fig3}.  (b) The same as (a) but for $U_K$=$-$120 MeV.
\protect\label{fig8}}
\end{figure}

\subsubsection{Dependence of the ground-state properties on $|S|$}
\label{subsubsec:3-2-2}
  
\ \ The effects of the nonlinear $\bar K-\bar K$ repulsive interaction in the chiral model reveal in saturation properties of other quantities. In Figs.~\ref{fig5} and \ref{fig8}, the chiral angle at $r=0$, $\theta^{(0)}$, and the baryon number density at $r=0$, $\rho_{\rm B}^{(0)}$ in the unit of the standard nuclear density $\rho_0$, are shown as functions of $|S|$ for the MKN with $A$=15, $Z$=8 in the case of $U_K$=$-$80 MeV and $-$120 MeV. The meaning of the curves is the same as those in Fig.~\ref{fig3}.
From comparison of the results in the chiral model with those in the lowest-order approximation in Figs.~\ref{fig5} and \ref{fig8}, one can see that the nonlinear $\bar K-\bar K$ interaction effects become significant as $|S|$ increases.
In the chiral model, the $\theta^{(0)}$ and $\rho_{\rm B}^{(0)}$ increase monotonically as $|S|$ increases up to $|S|\sim$ 8. In this respect, the $K^-$-bound state inside the MKN with a small (large) number of $|S|$ corresponds to a weak (fully-developed) kaon condensation in infinite matter. Due to the nonlinear effect, they are saturated for $|S|\gtrsim$ 8 for both $U_K$=$-$80 MeV and $-$120 MeV, while they continue to grow even for large $|S|$ in the lowest-order approximation. 
 For $|S|\gtrsim$ 8, the $\rho_{\rm B}^{(0)}$ is saturated to be $\rho_{\rm B}^{(0)}\sim$ 2.6 $\rho_0$ (4.0 $\rho_0$) for $U_K$ = $-$80 MeV ($-$120 MeV) in the chiral model. 

In the case of the MEM1 and MEM2, where the nonlinear $\bar K-\bar K$ terms coming from the nonlinear representation of the $K^-$ field are not taken into account, one can also see qualitative features similar to those in the lowest-order approximation: The $\theta^{(0)}$ and $\rho_{\rm B}^{(0)}$ increase monotonically with increase in $|S|$ and have larger values than those in the chiral model. However, the quantitative behaviors are dispersed among the results in the MEMs and those in the lowest-order approximation due to the additional terms (i) and (ii) in the MEMs. 

The saturation of the nuclear density distributions and $K^-$ field distribution with respect to the increase in $|S|$ results from the balance between the attractive energy coming from the $\bar K-N$ interaction and intermediate $\sigma$-exchange $N-N$ interaction,  and the repulsive energy from the $\bar K-\bar K$, the short-range $\omega$-exchange $N-N$ interactions, and the nucleon kinetic  energy. In particular, the nonlinear $\bar K-\bar K$ repulsion has a decisive role on the saturation mechanisms of these quantities, as seen in Figs.~\ref{fig5}$-$\ref{fig8}.  
On the other hand, in Ref.~\cite{gfgm07}, the saturation of the nuclear and $\bar K$ meson densities in $^{16}$O+$\kappa K^-$ was shown for the number of the $\bar K$ meson $\kappa\gtrsim$ 10, although the nonlinear $\bar K-\bar K$ repulsive interaction terms discussed in this paper are absent in their meson-exchange model. In their choice of the parameters, the coupling constant $g_{\omega N}$, which simulates the short-range $N-N$ repulsion, is taken to be a much larger value ($g_{\omega N}$ = 12.96) than ours ($g_{\omega N}$ = 8.72). Therefore, it is supposed that the short-range $N-N$ repulsion mainly overwhelms the $\bar K-N$ and intermediate $\sigma$-exchange $N-N$ attractions, keeping the nucleons apart as they would approach each other to avoid the short-range $N-N$ repulsion.

The chiral angle $\theta^{(0)}$ and the baryon number density $\rho_{\rm B}^{(0)}$ at the center of the nucleus depend quantitatively on the potential depth $U_K$ as shown in Figs.~\ref{fig5} and \ref{fig8}. 
For a larger value of $U_K$, for which the $\bar K N$ scalar attraction is larger, the $K^-$ mesons are located closer with each other to the center of the nucleus against the repulsion between $K^-$ mesons, and nucleons are attracted more to the center according to the distribution of the $K^-$. 
Therefore both the values of $\theta^{(0)}$ and $\rho_{\rm B}^{(0)}$ 
are larger for the larger value of $U_K$, which is valid in each case of the chiral model, MEMs and the lowest-order approximation. 

\begin{figure}[h]
\begin{minipage}[l]{0.50\textwidth}
\begin{center}
\includegraphics[height=.3\textheight]{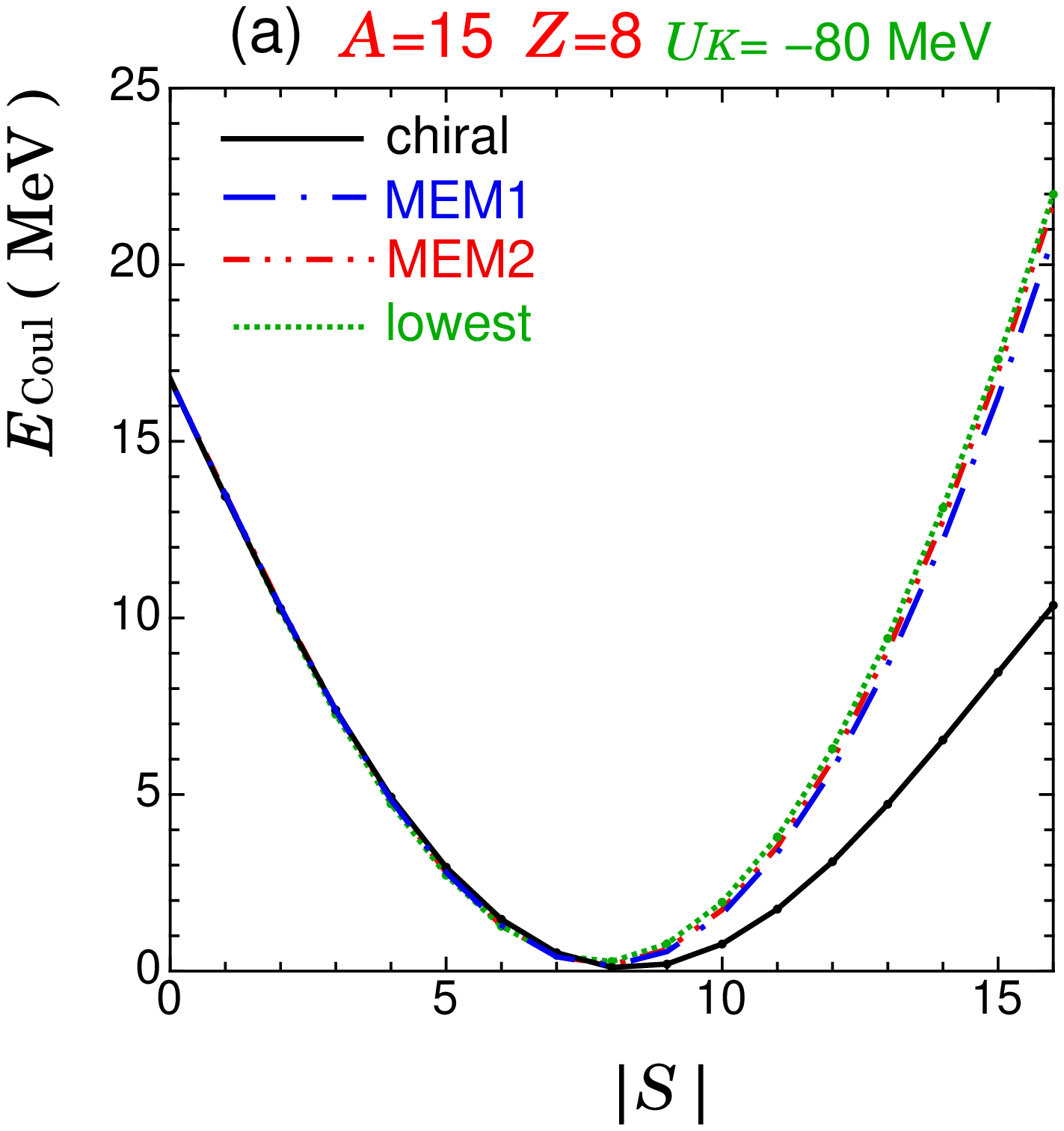}
\end{center}
\end{minipage}~
\begin{minipage}[r]{0.50\textwidth}
\begin{center}
\includegraphics[height=.3\textheight]{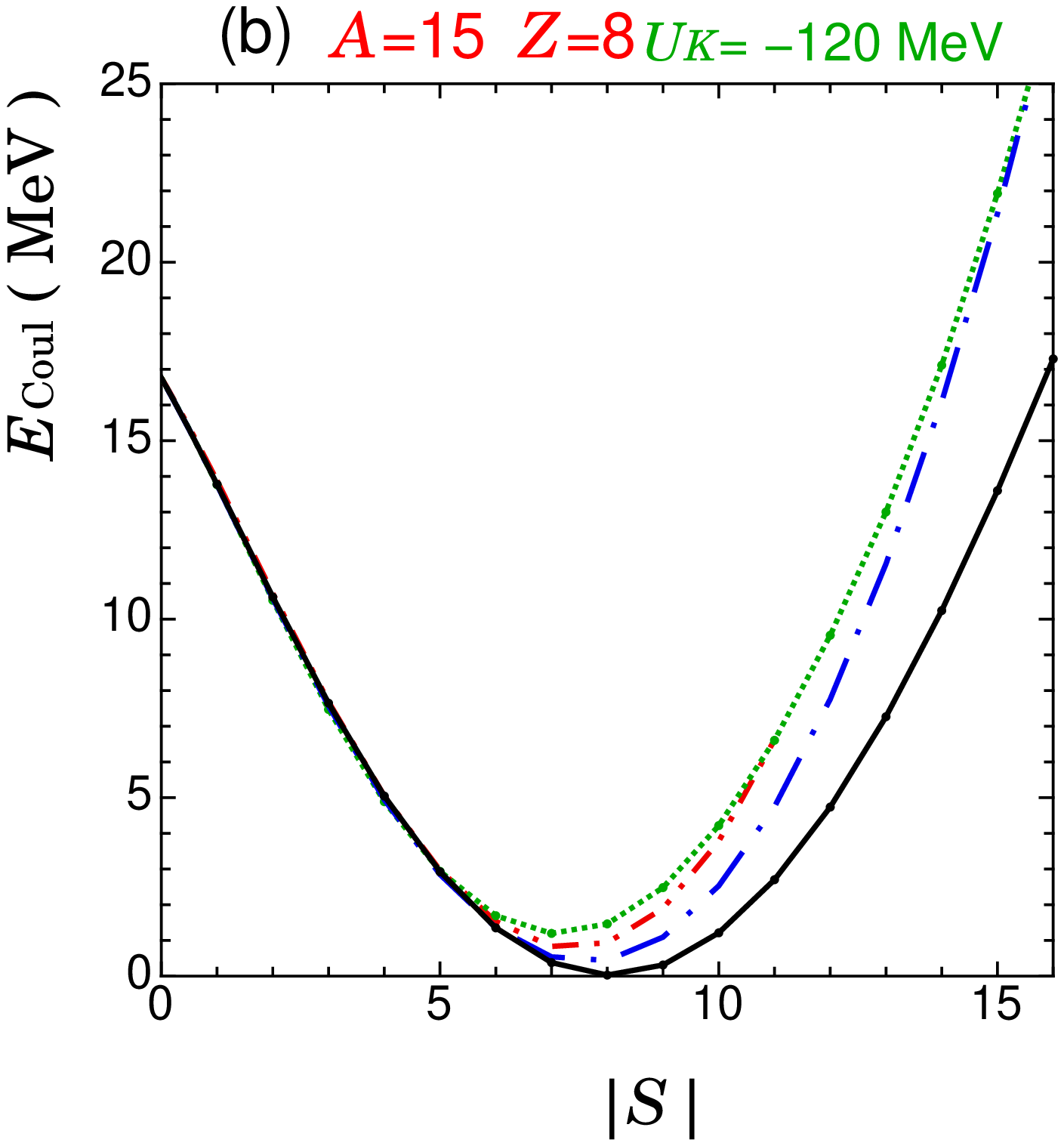}
\end{center}
\end{minipage}
\caption{(a) The Coulomb energy $E_{\rm Coul}$ of the MKN with $A$=15, $Z$=8 as functions of $|S|$ in the case of $U_K$=$-$80 MeV. The meaning of the curves is the same as in Fig.~\ref{fig3}.  (b) The same as (a) but for $U_K$=$-$120 MeV.
\protect\label{fig9}}
\end{figure}

\begin{figure}[!]
\begin{minipage}[l]{0.50\textwidth}
\begin{center}
\includegraphics[height=.3\textheight]{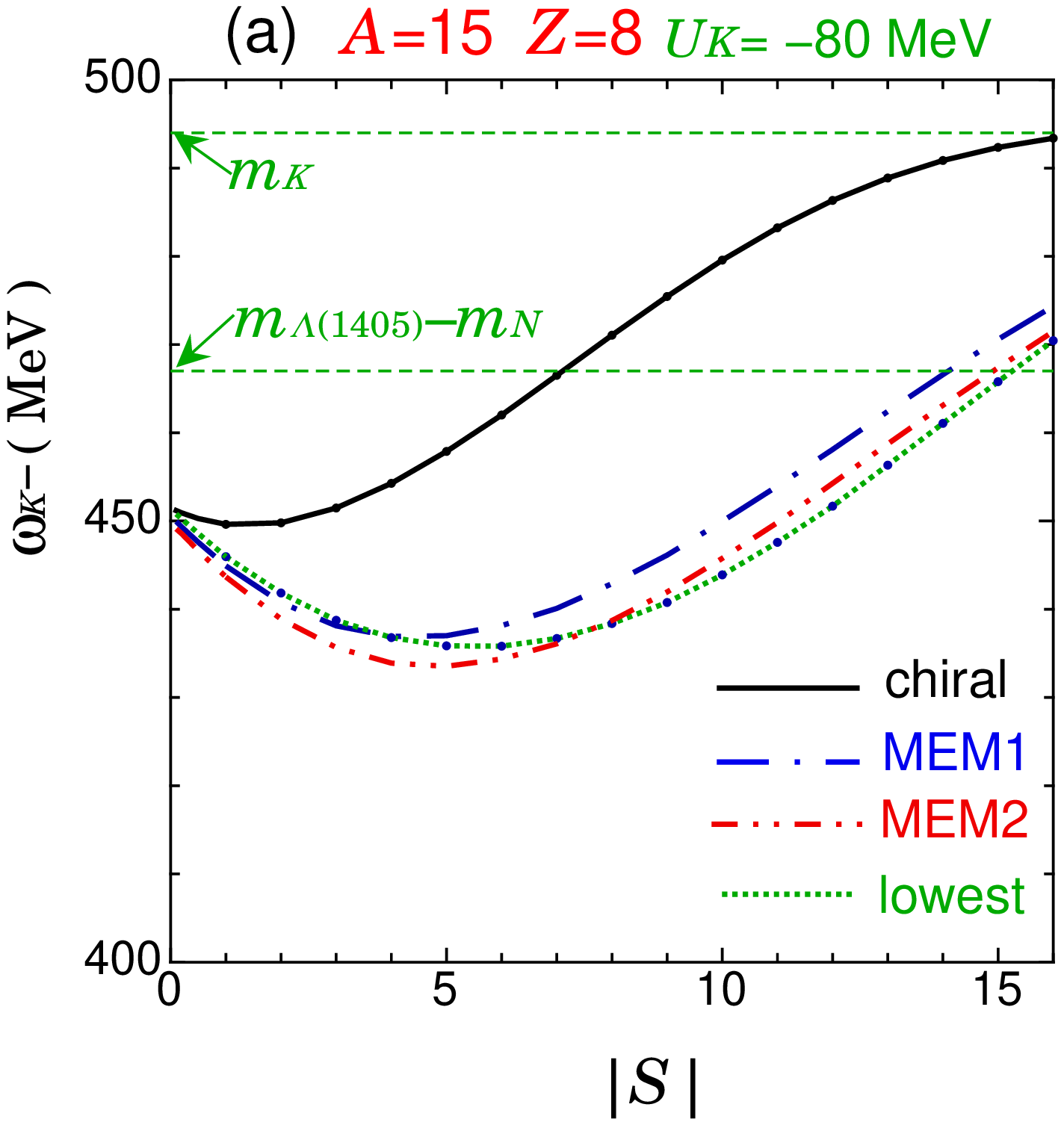}
\end{center}
\end{minipage}~
\begin{minipage}[r]{0.50\textwidth}
\begin{center}
\includegraphics[height=.3\textheight]{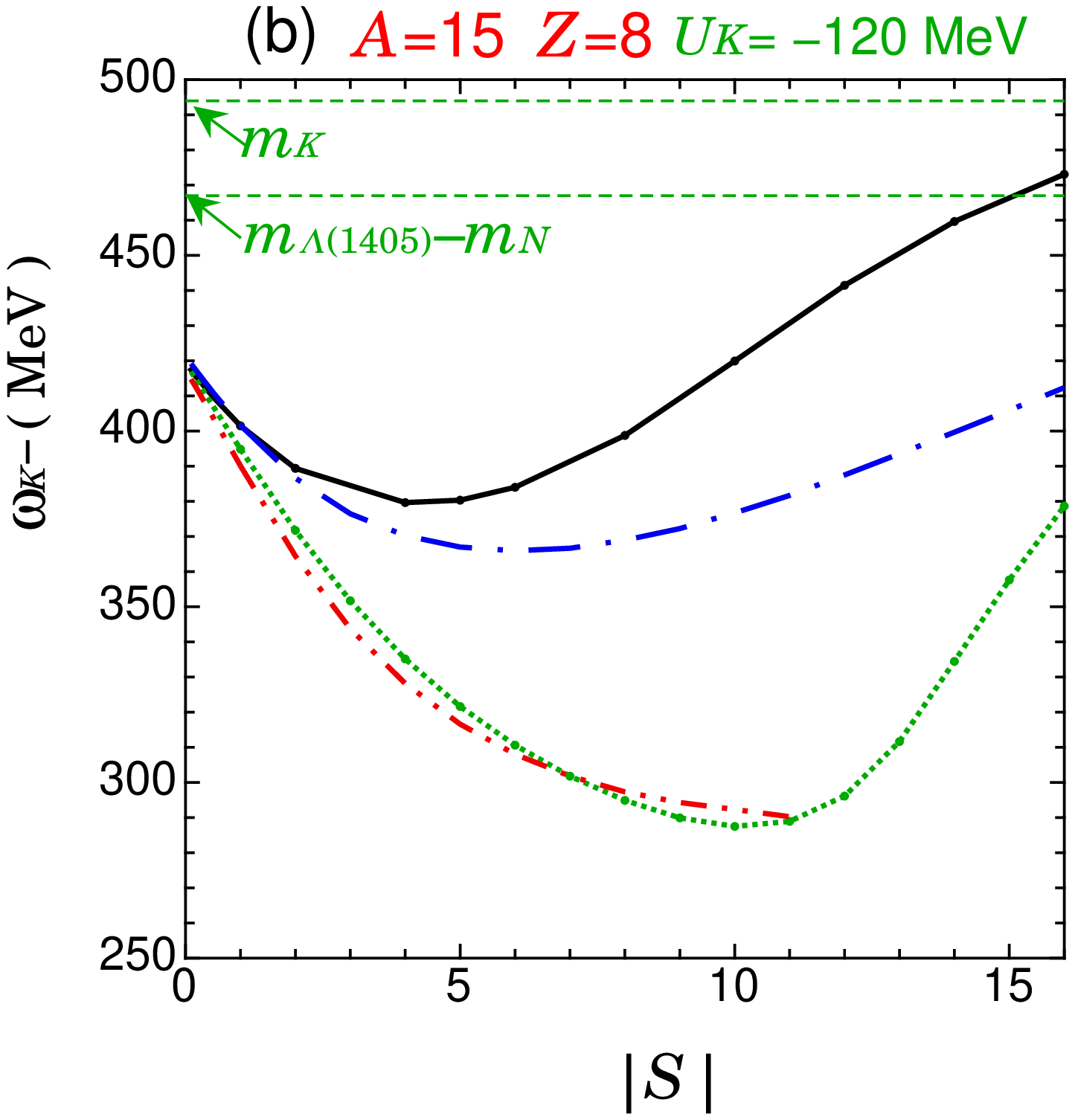}
\end{center}
\end{minipage}
\caption{(a) The lowest energy $\omega_{K^-}$ of the $K^-$ mesons embedded in the MKN with $A$=15, $Z$=8 as functions of $|S|$ in the case of $U_K$=$-$80 MeV. The meaning of the curves is the same as in Fig.~\ref{fig3}.  (b) The same as (a) but for $U_K$=$-$120 MeV. See the text for details. 
\protect\label{fig10}}
\end{figure}

\begin{figure}[!]
\begin{minipage}[l]{0.50\textwidth}
\begin{center}
\includegraphics[height=.3\textheight]{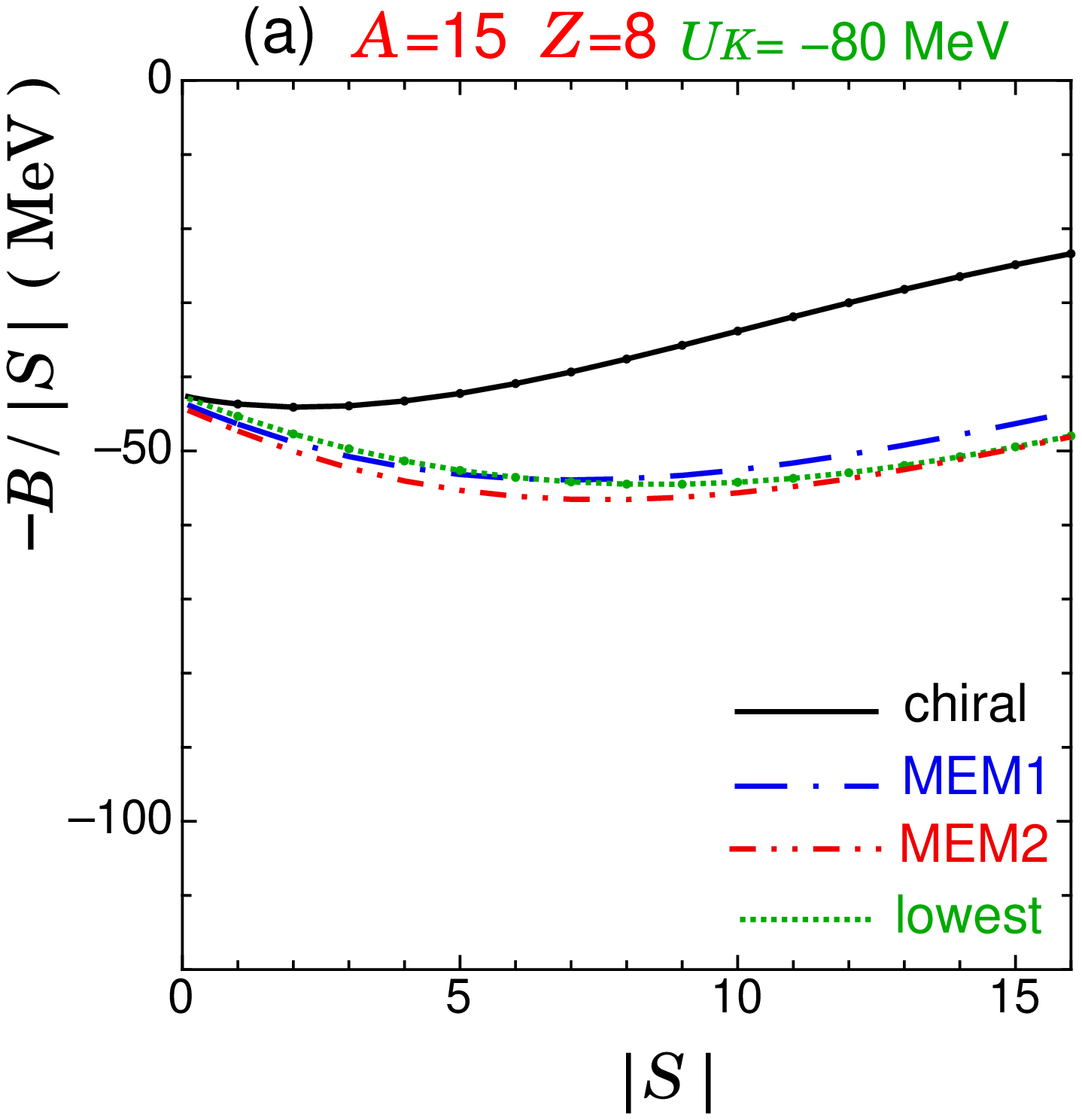}
\end{center}
\end{minipage}~
\begin{minipage}[r]{0.50\textwidth}
\begin{center}
\includegraphics[height=.3\textheight]{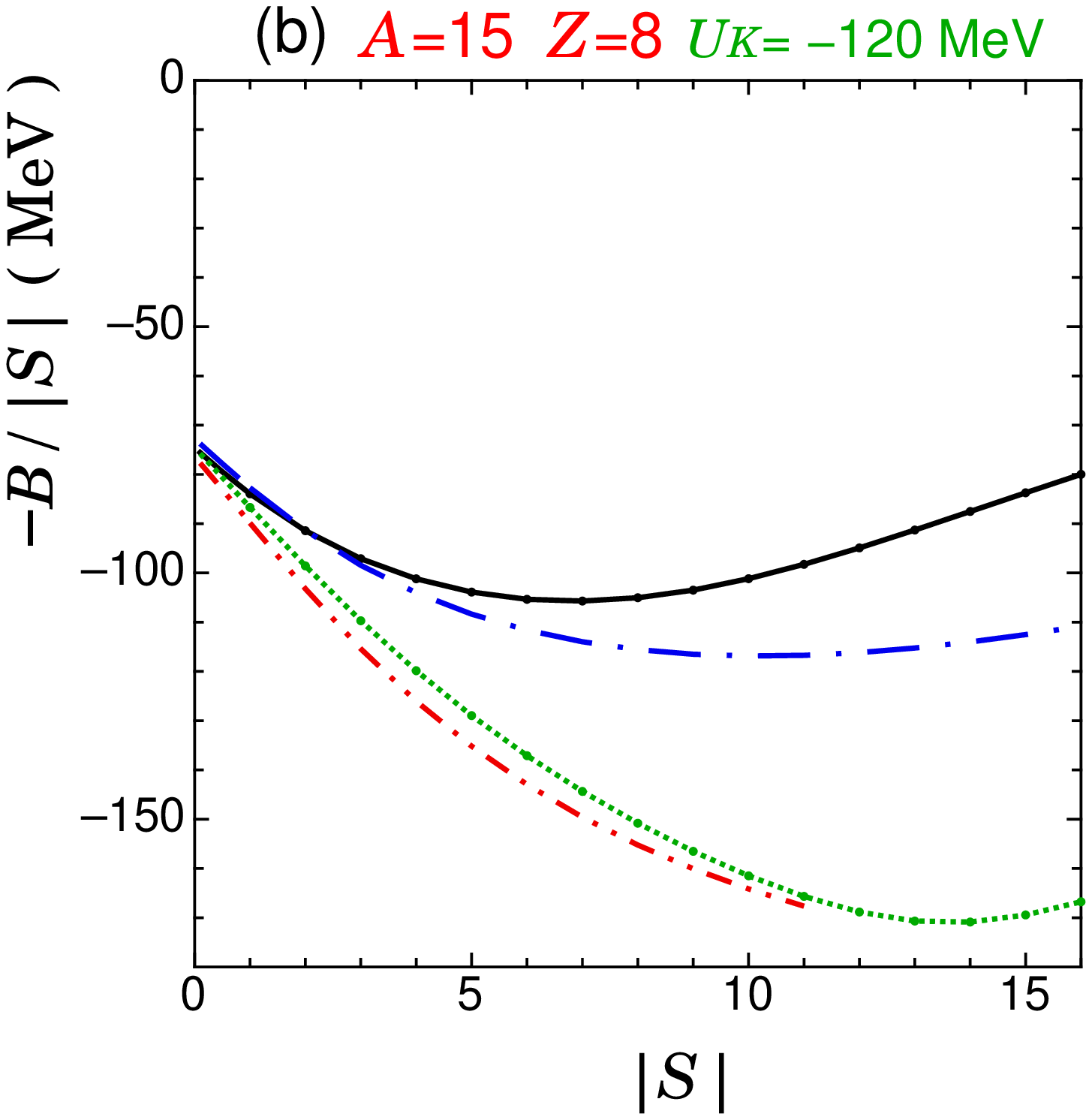}
\end{center}
\end{minipage}
\caption{(a) The sign-reversed binding energy per strangeness, $-B(A, Z, |S|)/|S|$, for the MKN with $A$=15, $Z$=8 as functions of $|S|$ in the case of $U_K$=$-$80 MeV. The meaning of the curves is the same as in Fig.~\ref{fig3}.  (b) The same as (a) but for $U_K$=$-$120 MeV.
\protect\label{fig11}}
\end{figure}

In Figs.~\ref{fig9}, \ref{fig10}, and \ref{fig11}, the Coulomb energy, $E_{\rm Coul}$ (=$\displaystyle \int d^3 r V_{\rm Coul}(r)$ ), the lowest energy $\omega_{K^-}$ of the $K^-$ meson, and the sign-reversed binding energy per strangeness $|S|$, $-B(A, Z, |S|)/|S|$, are shown as functions of $|S|$ for the MKN with $A$=15, $Z$=8 in the case of $U_K$=$-$80 MeV and $-$120 MeV. The binding energy for the MKN is defined by $B(A, Z, |S|)= E(A, Z, 0)+|S|m_K-E(A, Z, |S|)$, where $E(A, Z, |S|)$ is the ground state energy of the MKN measured from the sum of the free nucleon masses, $Am_N$.  

~
The Coulomb energy, $E_{\rm Coul}$, has a minimum around $|S|$ =$Z$ = 8, where the total electric charge vanishes for any models and for both cases of $U_K$.  From comparison of Fig.~\ref{fig9} with Figs.~\ref{fig10} and \ref{fig11}, the Coulomb energy is appreciably smaller than the $\omega_{K^-}$ and the binding energy $B(A, Z, |S|)$ at a fixed $|S|$. Thus the Coulomb energy has a minor contribution to the bulk properties of the MKN. In contrast, it is to be noted that the Coulomb effect has been shown to be important in a phase-equilibrated system through charge screening together with the surface energy effect\cite{vyt03,mtv06,mtec06}. 
For $|S|\gtrsim$ 8, the Coulomb energy increases monotonically with the increase in the magnitude of the negative total electric charge, $Z-|S|$. In the chiral model, the Coulomb energy has the smallest value at a given $|S|$ than the cases of the MEMs and that in the lowest-order approximation, as seen in Fig.~\ref{fig9}. This is explained as follows: As the number of the embedded $K^-$ mesons, $|S|$, gets larger, the $K^-$ mesons are more repelled from each other in the chiral model as compared with the other cases due to the existence of the nonlinear $\bar K-\bar K$ repulsive interaction terms, and so are the protons in accordance with the profile of the $K^-$ mesons. Thus localization of both the protons and $K^-$ mesons in the MKN are weak in the chiral model, which results in lowering the bulk Coulomb energy as compared with the other cases. 

The repulsive effects from the nonlinear $\bar K - \bar K$ interaction terms on both the $\omega_{K^-}$ and $ -B(A, Z, |S|)/|S|$ for large $|S|$ are evident from comparison of the results in the chiral model with those in the MEMs and the lowest-order approximation. 
The sign-reversed binding energy per strangeness is written as $ -B(A, Z, |S|)/|S|=[ E(A, Z, |S|)-E(A, Z, 0)]/|S|-m_K$.  For instance, in the MEM1, $ -B(A, Z, |S|)/|S|$ decreases as $|S|$ increases until $|S|\sim$ 8 ($|S|\sim$ 10) for $U_K$=$-$ 80 MeV ($U_K$=$-$ 120 MeV), which means that the energy difference, $E(A, Z, |S|)-E(A, Z, 0)$, decreases more rapidly than linearly in $|S|$. This is due to the fact that compression of the nucleus by embedding of the $K^-$ mesons enlarges attraction between the $K^-$ mesons and nucleons in addition to the case where the number of the $K^-$ mesons are simply added. At a sufficiently large $|S|$, $ -B(A, Z, |S|)/|S|$ turns to increase with $|S|$, since the $\bar K-\bar K$ repulsion  generated by the couplings between the $K^-$ and vector meson mean fields overcomes the $\bar K-N$ attraction. [See Sec.~\ref{subsec:kk} and Sec.~\ref{subsubsec:3-2-3}.] 

In the chiral model, the nonlinear $\bar K-\bar K$ repulsion further contributes to the increase in the ground state energy of the MKN. As a result,  
one has a maximum of the binding energy per strangeness, $(B/|S|)_{\rm max}$ = 44 MeV at $|S|$=2 for $U_K$=$-$80 MeV, and $(B/|S|)_{\rm max}$ = 106 MeV at $|S|$=7 for $U_K$=$-$120 MeV in the chiral model.

From comparison with the cases of $U_K$=$-$ 80 MeV and $-$ 120 MeV in Figs.~\ref{fig10} and \ref{fig11}, one can naturally see that the deeper $\bar K N$ attractive potential leads to the lower energy for the $\omega_{K^-}$ and more binding energy $B(A, Z, |S|)/|S|$ at a given $|S|$. For $|S|\sim A$, the repulsive effects compensate for the $\bar K-N$ attractions, so that the $\omega_{K^-}$ tends to the free kaon mass, $m_K$, for both $U_K$=$-$80 MeV and $-$120 MeV. It is to be noted that the $\omega_{K^-}$ in the chiral model exceeds the mass difference of the $\Lambda$(1405) ($\Lambda^\ast$) and nucleon (=467 MeV) for $|S|\gtrsim$7 ($|S|\gtrsim$15) in the case of $U_K$=$-$80 MeV ($U_K$=$-$120 MeV). 

In this paper we do not take care much about the $\Lambda^\ast$. The neglect of the $\Lambda^\ast$ may be justified for the deeply bound kaons, as in the case of $U_K=-120$MeV or $|S|\lesssim 6$ for $U_K=-80$MeV.
For one or two $K^-$ mesons they are deeply bound inside nuclei and their energy is well below the resonance region, 
$\omega_{K^-}\simeq m_{\Lambda(1405)}-m_N$.
Since $\omega_{K^-}$ increases due to the $\bar K -\bar K$ interaction as the number of the $K^-$ increases, it enters into the
resonance region. Then kaons are strongly coupled with
$\Lambda^\ast$-hole state. In such a case it would be necessary to take into account the $\Lambda^\ast$ excitation explicitly. It is expected that the $K^-$ state couples with the $\Lambda^\ast$-hole state, resulting in level-crossings between the real $K^-$ state and the $\Lambda^\ast$-hole state, which may modify the microscopic properties of the MKN. Moreover, it may have some implications on kaon condensation in compact stars.

\subsubsection{$\bar K-\bar K$ interaction from coupling between the $K^-$ and meson mean-fields}
\label{subsubsec:3-2-3}

\ \ One can see that both $\omega_{K^-}$ and $-B(A, Z, |S|)/|S|$ are pushed up at large $|S|$ even in the case of the MEMs and the lowest-order approximation, where the nonlinear $\bar K-\bar K$ terms coming from the nonlinear representation of the $K^-$ field are absent. 
Such an additional repulsion can be interpreted with recourse to the $\bar K-\bar K$ interaction generated by the couplings between the $K^-$ and scalar and vector mean fields, which is discussed in Sec.~\ref{subsec:kk}.  For instance, we look into detail this repulsive effect on the $\omega_{K^-}$ within the MEM1. We formally express the $ \omega_{K^-}$ as $\displaystyle \omega_{K^-}=V_{\rm Coul}(r)-X_0(r)+\lbrack {m_K^\ast}_{\rm ,MEM1}^2(r)-\nabla^2\theta(r)/\theta(r)\rbrack^{1/2}$ by the use of the equation of motion for $\theta$, Eq.~(\ref{eq:lin4}). 
The main contribution to the $\omega_{K^-}$ comes from 
${m_K^\ast}_{\rm ,MEM1}(r)$ and $X_0(r)$, both of which are represented as Eqs.~(\ref{eq:elims}) and (\ref{eq:elimv}), respectively,  with the use of the equations of motion for the scalar and vector mean fields, Eqs.~(\ref{eq:lin1})$-$(\ref{eq:lin3}).  In Sec.~\ref{subsec:kk}, it has been shown that the third term in the bracket on the r.~h.~s. of Eq.~(\ref{eq:elims}) is generated from the $\bar K-\bar K$ interaction mediated by the $\sigma$-meson exchange, and it contributes to a decrease of ${m_K^\ast}_{\rm ,MEM1}(r)$. Likewise, the third term in the bracket on the r.~h.~s. of Eq.~(\ref{eq:elimv}) is generated from the $\bar K-\bar K$ interaction mediated by the $\omega$ and $\rho$-mesons exchange, and it contributes to a decrease of $X_0(r)$. 

\begin{figure}[!]
\begin{minipage}[l]{0.50\textwidth}
\begin{center}
\includegraphics[height=.3\textheight]{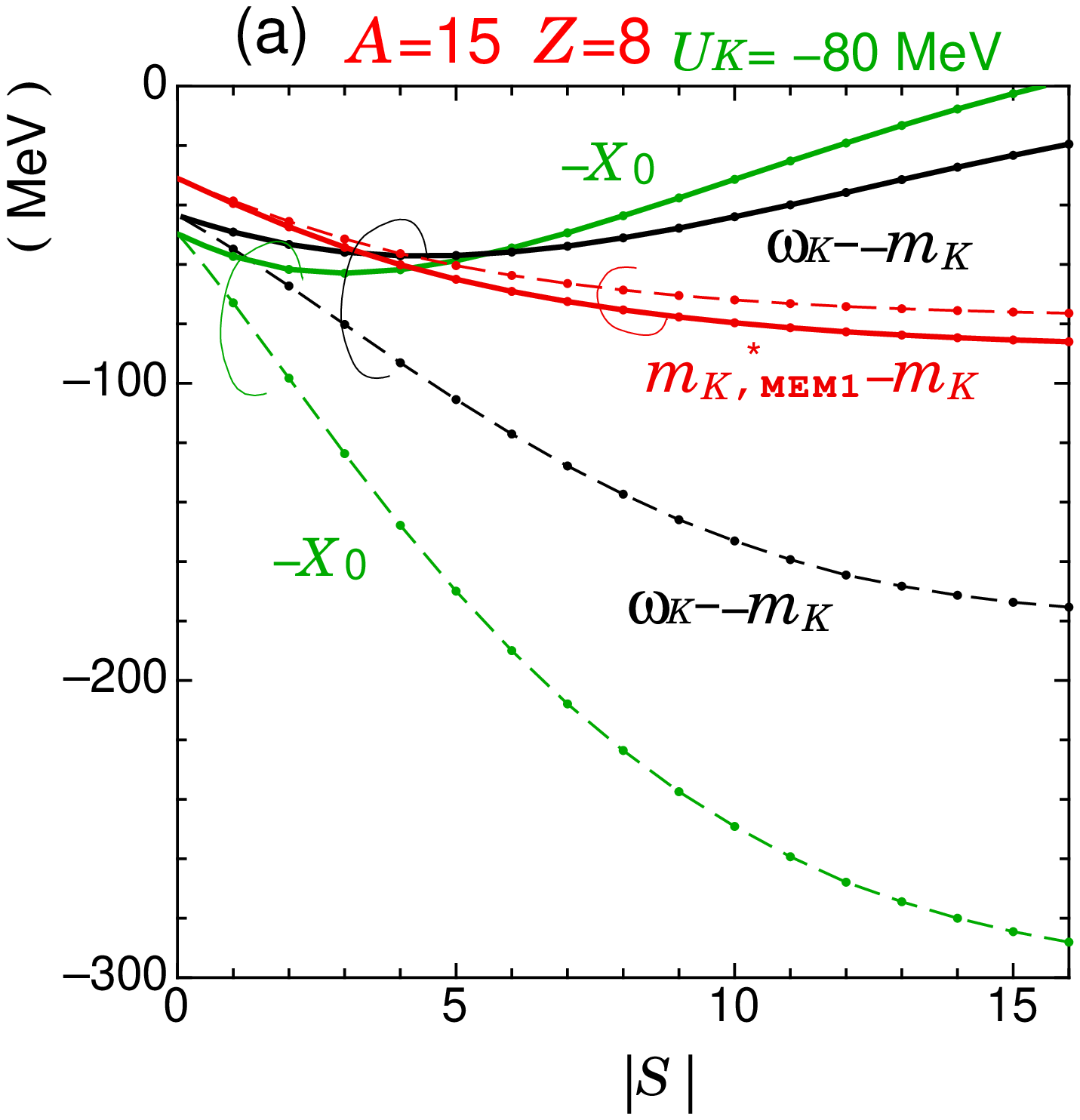}
\end{center}
\end{minipage}~
\begin{minipage}[r]{0.50\textwidth}
\begin{center}
\includegraphics[height=.3\textheight]{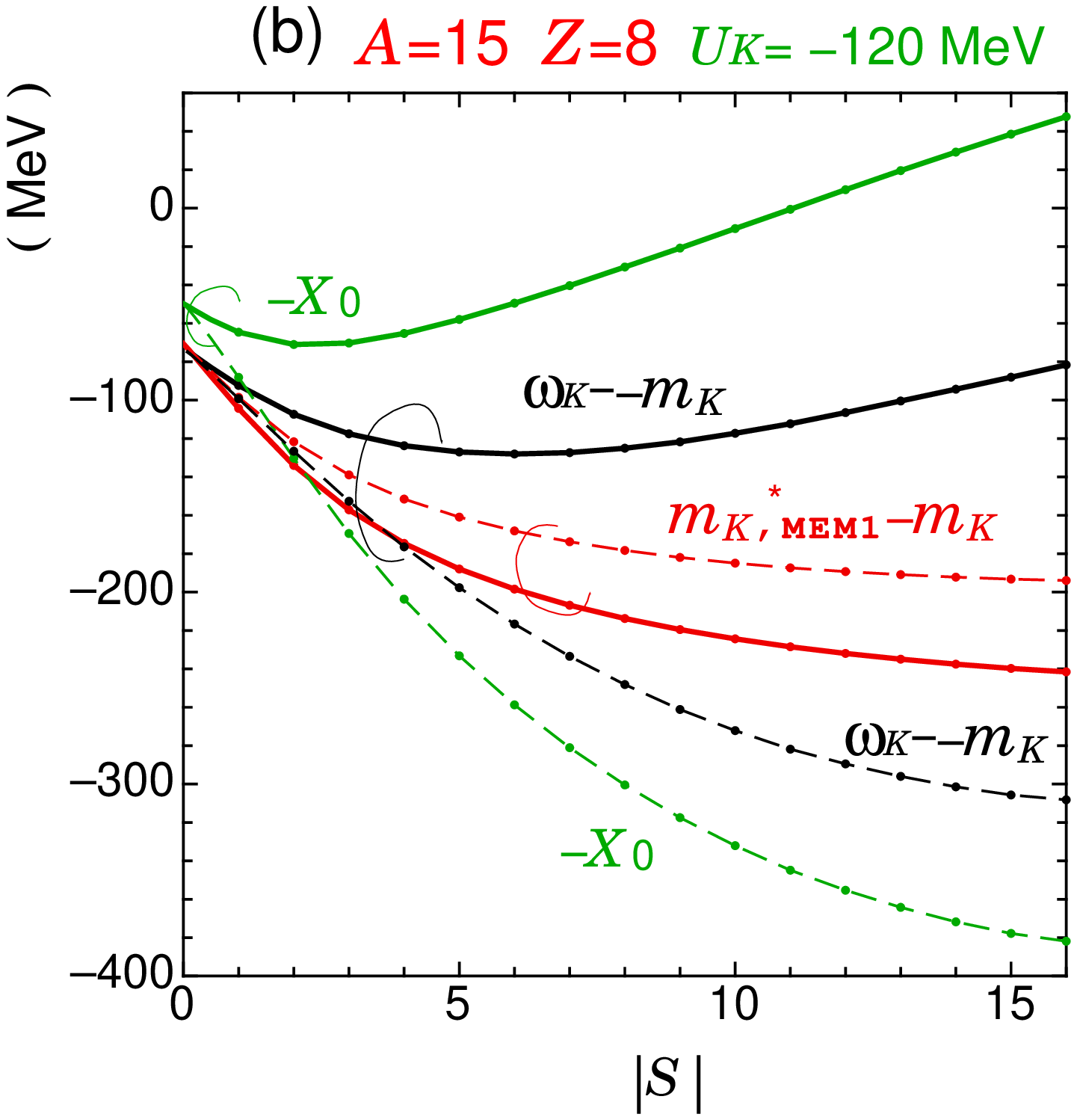}
\end{center}
\end{minipage}
\caption{(a) The lowest $K^-$ energy measured from the free kaon mass, $\omega_{K^-}-m_K$, the effective mass for the $K^-$ meson  measured from the free kaon mass, ${m_K^\ast}_{\rm ,MEM1}^{(0)}-m_K$, and $-X_0^{(0)}$ in the MEM1 as functions of $|S|$ for $U_K$=$-$80 MeV (the solid lines). The superscripts $(0)$ attached to these quantities stand for those estimated at $r=0$. Also shown by the dashed lines are these quantities obtained with the kaon-coupling terms in the equations of motion for the $\sigma$, $\omega$, and $\rho$ fields being turned off, while the density profiles of $\rho_p(r)$ and $\rho_n(r)$ are artificially kept fixed to be the same as those obtained by the full expressions [Eqs.~(\ref{eq:lin1})$-$(\ref{eq:lin3})]. 
 (b) The same as (a) but for $U_K$=$-$120 MeV.
\protect\label{fig12}}
\end{figure}

In order to see the effects of these $K^-$-scalar and vector mesons coupling terms quantitatively, 
we evaluate the $\omega_{K^-}$ in terms of ${m_K^\ast}_{\rm ,MEM1}(r)$ and $X_0(r)$ at $r=0$. 
We show, in Fig.~\ref{fig12}(a) [Fig.~\ref{fig12}(b)], the lowest $K^-$ energy measured from the free kaon mass, $\omega_{K^-}-m_K$, the effective mass of the $K^-$ measured from the free kaon mass, ${m_K^\ast}_{\rm ,MEM1}^{(0)}-m_K$, and $-X_0^{(0)}$ in the MEM1 as functions of $|S|$ for $U_K$=$-$80 MeV [$U_K$=$-$120 MeV] by the solid lines. [The superscripts $(0)$ attached to these quantities stand for those estimated at $r=0$.] We also show, by the dashed lines, these quantities which we obtained by turning off  
the kaon-coupling terms in the equations of motion for the $\sigma$, $\omega$, and $\rho$ fields, while the density profiles of $\rho_p(r)$ and $\rho_n(r)$ are artificially kept fixed to be the same as those obtained by the full expressions [Eqs.~(\ref{eq:lin1})$-$(\ref{eq:lin3})]. 
One finds that ${m_K^\ast}_{\rm ,MEM1}^{(0)}-m_K$ decreases with increase in $|S|$ for both cases where the coupling term between the $K^-$ and scalar meson exists (the solid lines) and where it is turned off (the dashed lines). It is shown from comparison between the solid and dashed lines that the effective mass of the $K^-$ is further reduced as a result of the coupling between the $K^-$ and scalar meson with increase in $|S|$, leading to reduce the $\omega_{K^-}$ further.
On the other hand, $-X_0^{(0)}$ decreases in magnitude with $|S|$ at large $|S|$ (solid lines), while it steadily increases in magnitude when the coupling terms between the $K^-$ and vector mesons are turned off (dashed lines). In addition, it is shown from comparison between the solid and dashed lines that the coupling terms between the $K^-$ and vector mesons drastically suppress the $-X_0^{(0)}$ with increase in $|S|$, leading to push up the value of $\omega_{K^-}$. As a result, the coupling terms between the $K^-$ and vector mesons mainly serve to increase the $\omega_{K^-}$ at large $|S|$. 
~
\begin{figure}[!]
\begin{minipage}[l]{0.50\textwidth}
\begin{center}
\includegraphics[height=.3\textheight]{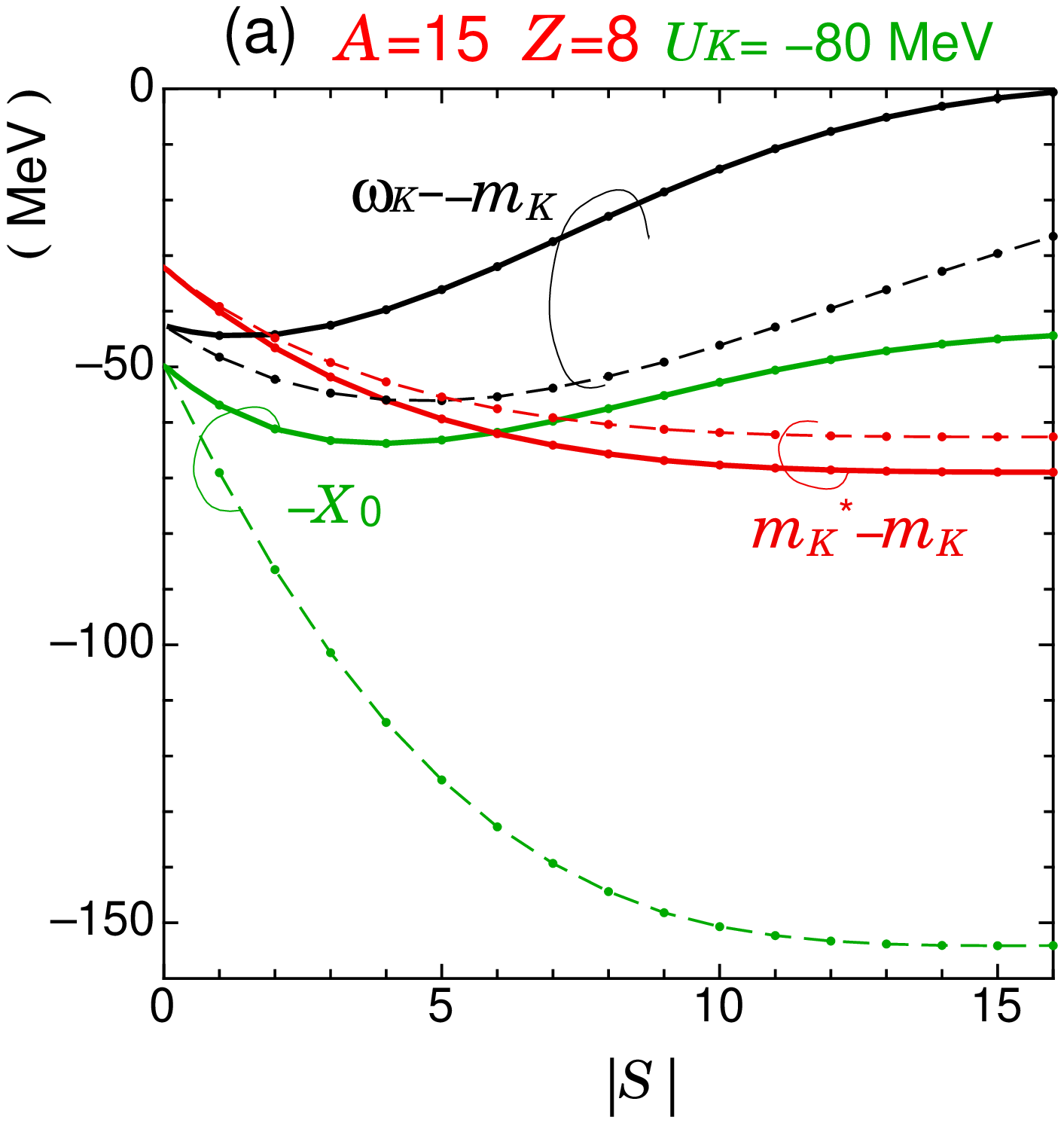}
\end{center}
\end{minipage}~
\begin{minipage}[r]{0.50\textwidth}
\begin{center}
\includegraphics[height=.3\textheight]{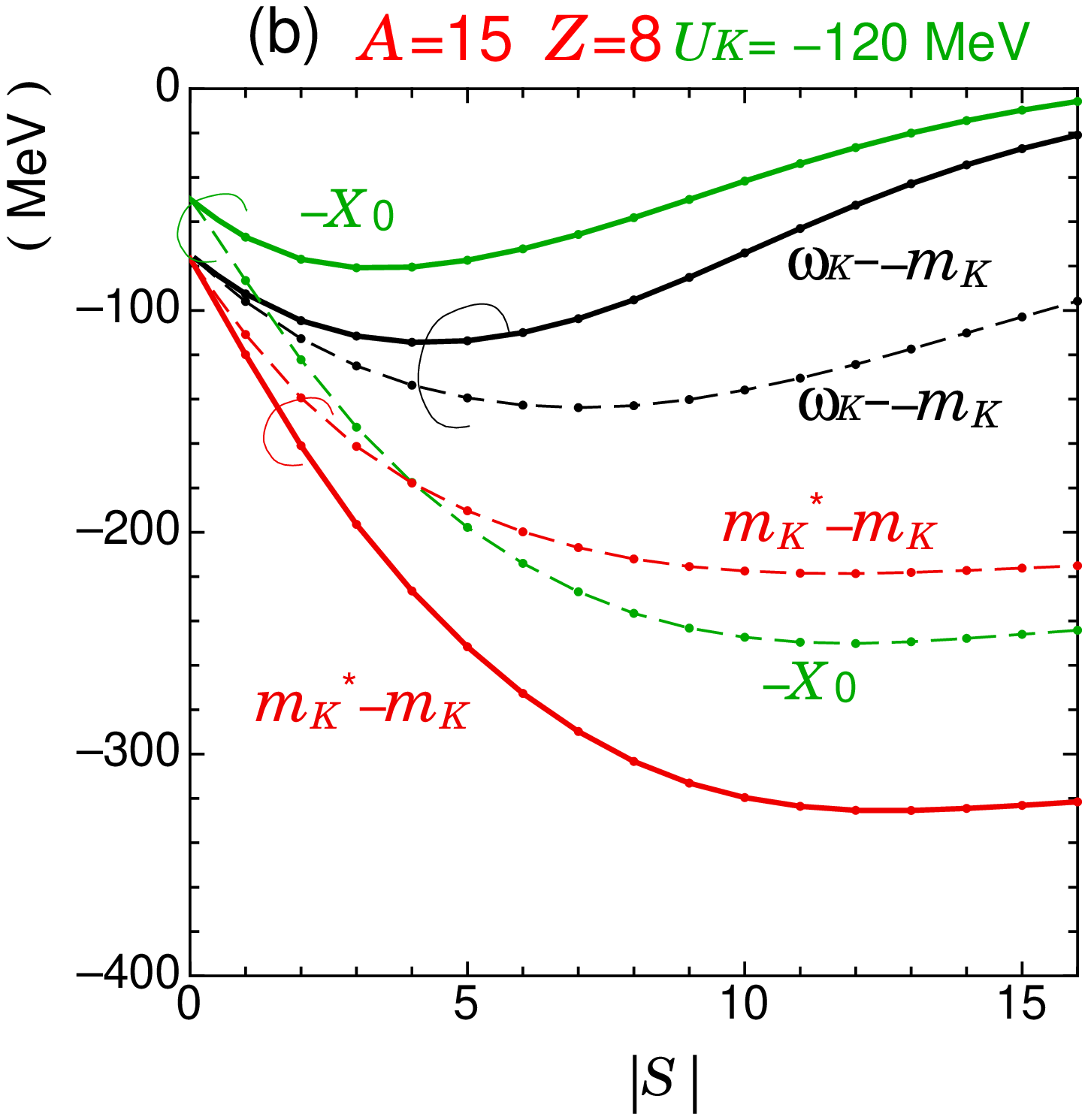}
\end{center}
\end{minipage}
\caption{The same as Fig.~\ref{fig12} but for the results obtained in the chiral model.
\protect\label{fig13}}
\end{figure}

In Fig.~\ref{fig13}~(a) [Fig.~\ref{fig13}~(b)], the quantities, $\omega_{K^-}-m_K$, ${m_K^\ast}^{(0)}-m_K$, and $-X_0^{(0)}$ are shown as function of $|S|$ in the chiral model for $U_K$=$-$80 MeV [$U_K$=$-$120 MeV]. The same mechanism of the $\bar K-\bar K$ repulsive interaction terms through the $K^-$-scalar and vector mesons-coupling terms works also in the chiral model. For both the chiral model and MEMs, the $K^-$-vector meson-coupling terms suppress the $\bar K- N$ attractive vector interaction at large $|S|$, while the $K^-$-scalar meson-coupling term slightly enhances the $\bar K- N$ attractive scalar interaction. Since the former effect is much larger than the latter, $\omega_{K^-}$ turns to increase at large $|S|$.

\subsection{Implications for experiments}
\label{subsec:implications}

\ \ Following the results in the previous sections,  we take up a reference nucleus $^{15}_{\ 8}$O ($A$=15, $Z$=8) with no trapped $K^-$ meson to discuss a possible observation of the MKN in experiments. We base our discussion on the chiral model. 

First we consider the $\bar K$ nuclei with single $K^-$ meson ($|S|$ =1), for which the experimental searches have been done most extensively\cite{i03,k05,a05}. The central baryon number density $\rho_{\rm B}^{(0)}$ and the binding energy $B(A, Z, |S|)$ are read from the results in the chiral model (Figs.~\ref{fig8} and \ref{fig11}): In the case of $U_K$=$-$80 MeV, ($\rho_{\rm B}^{(0)}$, $B$)=(1.4$\rho_0$, 44 MeV), and in the case of $U_K$=$-$120 MeV, ($\rho_{\rm B}^{(0)}$,$B$)=(1.7$\rho_0$, 84 MeV).
The $\bar K$-nucleus interaction has been studied by the in-flight $^{16}$O ($K^-$, $N$) and $^{12}$C ($K^-$, $N$) reactions\cite{k05}. The deep $\bar K$-nucleus potential of around $-$ 200 MeV has been derived for the binding energy of the ground state of the kaonic nuclei around 100 MeV from the analysis of the missing mass spectra of these reactions. In Ref.~\cite{ynh06}, the formation spectra for these reactions have been obtained theoretically, but no clear peak structure for the kaonic nuclei was found, which agrees with the experimental data\cite{k05}. Nevertheless, according to our result, the central baryon number density of the $\bar K$ nuclei shows a definite increase from $\rho_0$ although it does not reach 2$\rho_0$, and one expects a sizable binding energy, in particular, in the case of $U_K$=$-$120 MeV. These quantities will serve as additional observables indicating the formation of deeply bound kaonic nuclear states other than the spectra of the reactions. 

Second, we are concerned with the MKN with several numbers of the trapped $K^-$ mesons. It has been proposed that double and/or multiple antikaon clusters may be identified in fragments after the relativistic heavy-ion collisions by way of invariant mass spectroscopy for decaying particles of $\bar K$ clusters\cite{yda04}. 
New facilities such as the Facility for Antiproton and Ion Research (FAIR) at GSI are expected to contribute to a  possible formation of the MKN. As another formation scenario of the MKN, fusion processes of the single or double $\bar K$ nuclei may help sequentially increase the number of the trapped $K^-$ mesons in the nucleus. To be more realistic, one has to consider the quantities such as fusion cross sections which are directly accessible to experimental data. 

Highly dense object would be obtained if the number of the trapped $K^-$ mesons can be as large as $|S|=O(A)$.
In our results, the central nuclear density reaches $\rho_{\rm B}^{(0)}\sim$ 2.6$\rho_0$ (4.0$\rho_0$) for $U_K$=$-$80 MeV ($U_K$=$-$120 MeV) at a large value of $|S|$ ($\gtrsim$ 8). The binding energy per strangeness has a maximum value, $B(A, Z, |S|) /|S|$$\sim$ 40 MeV ($\sim$100 MeV)  for $U_K$=$-$80 MeV ($U_K$=$-$120 MeV), as discussed in Sec.~\ref{subsubsec:3-2-2} (Fig.~\ref{fig11}). Thus the formation of the MKN accompanies a large energy release through  emission of particles and radiation, especially in the case of $U_K$=$-$120 MeV. At a large value of $|S|$, $\bar K-\bar K$ repulsion has a substantial effect on the reduction of the binding energy, $B(A, Z, |S|) /|S|$. 
Even for the double $\bar K$ nuclei ($|S|$=2), one can already see the nonlinear $\bar K-\bar K$ repulsive effect on the binding energy, although it is quantitatively small, from comparison of the result of the chiral model with that in the lowest-order approximation (Fig.~\ref{fig11}).  
Therefore, we should carefully separate the $\bar K-\bar K$ interaction to extract information on the $\bar K-N$ interaction in such a dense medium as the MKN with large $|S|$. 

The ground state energy per strangeness for the MKN, $E(A, Z, |S|)/|S|$, is written as \\ $ E(A, Z, |S|)/|S| =-B(A, Z, |S|)/|S|+E(A, Z, 0)/|S|+m_K$.  Since $E(A, Z, 0)$=$-$104 MeV and  $-B/|S|\gtrsim$ $-$40 MeV ($-B/|S|\gtrsim$ $-$100 MeV) for $U_K$=$-$ 80 MeV ($U_K$=$-$ 120 MeV) in our result, $E(A, Z, |S|) / |S|$ exceeds the mass-difference between the lightest hyperon $\Lambda$(1116) and nucleon, $m_{\Lambda({\rm 1116})}-m_N $(= 176 MeV), for all the cases of $|S|$ and $U_K$. Hence the MKN is not absolutely stable, and it would decay into mesic or nonmesic final states through strong processes such as $\bar K NN\rightarrow \Lambda N$.  Similar result has been obtained in Ref.~\cite{gfgm07}, where it has been concluded that $\bar K$ mesons do not condense.
 Note that, in Ref.~\cite{gfgm07}, they mean ``kaon condensation'' in the laboratory by a stable self-bound matter that might be obtained by conversion from multi-strange hypernuclei by the strong process, $\Lambda\rightarrow \bar K + N$. 

As shown in Sec.~\ref{subsubsec:3-2-1}, 
a neutron skin appears for the MKN on the tails of the nucleon density profiles for a large $|S|$. The neutron skin structure may be confirmed by measurement of the RMS radii of the proton and neutron, which may in turn provide us with information of the isospin dependence of the $\bar K-N$ interaction.

\section{Summary and concluding remarks}
\label{sec:summary}

\ \ We have investigated the structure of the multi-antikaonic nuclei (MKN) in the relativistic mean-field theory by taking into account kaon dynamics on the basis of chiral symmetry. Effects of the nonlinear $\bar K-\bar K$ interaction inherent in the chiral model on the properties of the MKN have been clarified. For the nucleus with mass number $A$=15 and atomic number $Z$=8, we have obtained density distributions of the protons, neutrons, and $K^-$ mesons by systematically changing the number of the embedded $K^-$. It has been shown that the distributions of the nucleons and $K^-$ mesons tend to become uniform in the MKN due to the nonlinear $\bar K-\bar K$ repulsive interaction. The nonlinear $\bar K-\bar K$ repulsive effects become important for a large $|S|$ as follows: The distributions of the nucleons, the $K^-$ field (the chiral angle $\theta^{(0)}$) and the baryon number density $\rho_{\rm B}^{(0)}$ at the center of the nucleus are saturated for $|S|\geq$ 8. Furthermore, the saturated values of the $\theta^{(0)}$ and $\rho_{\rm B}^{(0)}$ are much reduced in comparison with those in the MEMs and in the lowest-order approximation. From the results on the structure of the MKN, it is summarized that the MKN with a small $|S|$ (with a large $|S|$) corresponds to weak kaon condensation (fully-developed kaon condensation) in infinite matter, although the nuclear density in the central region of the MKN is saturated for a large $|S|$. 

We have also seen that the $\bar K-\bar K$ interaction through the couplings between the $K^-$ and vector mesons gives a repulsive effect as well as the nonlinear terms of 
the contact interaction included in the original chiral Lagrangian.
It would be then interesting to study implications of the repulsive 
$\bar K-\bar K$ interaction through the meson-exchange on
kaon condensation in dense nucleon matter, since it is well known from the previous studies that 
the EOS is remarkably softened within the models based on chiral symmetry, where the $\bar K-N$ and $\bar K-\bar K$ interactions are furnished by the contact interaction, but without meson-exchange  between $\bar K$ mesons and nucleons\cite{kn86,mtt93,lbm95,fmmt96,tpl94}. The increase of the $\bar K-\bar K$ repulsive effect with increase in $|S|$ in the case of  the MKN suggests that the $\bar K-\bar K$ repulsion also plays an important role on the EOS at high baryon densities in the fully-developed kaon-condensed phase which may be realized in neutron stars. Such a repulsive effect may keep the EOS from getting too soft, and it may help make the neutron star mass consistent with observations\cite{l07}. 
In the extreme case, the mixed phase develops at very low densities.
It is instructive to reconsider kaon condensation in neutron stars from a viewpoint of the MKN. 
One may extract some information about $\bar K-\bar K$ interaction in nuclear medium by studying MKN at the new
facilities, J-PARC at JAEA/KEK and FAIR at GSI.

It has been shown that the effective $\bar K-\bar K$ repulsion leads to increase the lowest $K^-$ energy $\omega_{K^-}$ and to reduce the binding energy of the MKN at a large number of the embedded $K^-$ mesons $|S|$. Then one faces a situation where the lowest $K^-$ energy enters into the resonance region, $\omega_{K^-}\simeq m_{\Lambda (1405)}-m_N$ = 467 MeV. In such a case, the strong coupling of the $K^-$ meson with the $\Lambda(1405)$-hole state occurs. To treat the coupling properly, we must explicitly introduce the $\Lambda(1405)$ in our framework from the beginning and solve the coupled channel problem. Such work is now in
progress and we will report some results elsewhere. 

We did not take into account effects mediated by hyperons  for properties of the MKN. Channel-coupling effects through inelastic processes, $\bar KN\rightarrow \pi\Lambda, \pi\Sigma$, should be relevant to energy and decay width of the MKN. A part of the channel coupling effects can be incorporated by considering baryons as quasi-particles that are superposition of the nucleons and hyperons through the $p$-wave $\bar KNY$ couplings\cite{m02}. In addition, hyperon-mixing in the ground state of the MKN should lead to more bound and denser nuclei, which may be stable against strong processes, decaying only through weak interaction processes\cite{m08a,m08b}.

In our model, spherical symmetry has been assumed for the nucleon and $K^-$ meson profiles for the ground state of the MKN, and microscopic properties for the nuclear configuration such as shell structure and clustering structure have been ignored.
The embedded $K^-$ mesons may affect such microscopic structures of the nucleus. For instance, the strong attraction between the $K^-$ mesons distributed near the center and nucleons may stabilize the MKN against a rapid rotation. If the MKN is formed with a high angular momentum in a process of nuclear collisions , it may appear with a superdeformation of the excited nucleus. Such a deformation in turn changes the nuclear mean-field potential and thus modifies the shell structure for nucleons self-consistently. 
  
\section*{Acknowledgments}
\ \ This work is supported in part by the Grant-in-Aid for Scientific Research (Nos.~18540288, 19540313, and 20028009).

\end{document}